\documentclass[aps,prd,reprint,superscriptaddress,floatfix]{revtex4-2}

\usepackage[T1]{fontenc}
\usepackage{amsmath,amssymb}
\usepackage{bm}

\usepackage{microtype}
\usepackage{tikz}
\usetikzlibrary{positioning,arrows.meta,fit,backgrounds}
\usepackage{graphicx}
\usepackage{placeins}
\usepackage[colorlinks=true,allcolors=blue]{hyperref}

\newcommand{\EnWDF}{\ensuremath{\rho_{\mathrm{WDF}}}}

\begin{document}

\title{The Wavelet Detection Filter: a real time unmodeled pipeline
for gravitational wave transients, ranking coincidences
with a graph neural network}

\author{Elena Cuoco}
\email{elena.cuoco@unibo.it}
\affiliation{Physics and Astronomy Department (DIFA), Alma Mater Studiorum ---
Universit\`a di Bologna, Italy}
\affiliation{INFN, Sezione di Bologna, Viale C. Berti Pichat 6/2, 40126
Bologna, Italy}

\date{September 11, 2026}
\begin{abstract}
Accurate waveform models are unavailable for many potential sources of gravitational wave transients, 
motivating searches that do not assume a predefined template family. 
The Wavelet Detection Filter (WDF) is an unmodeled pipeline that identifies excess energy in the wavelet 
coefficients of whitened detector data. 
Here, we develop WDF into a low latency pipeline for both detection and
reconstruction and this paper describes and validates that pipeline. 
We use a zero phase streaming autoregressive whitening
filter that preserves transient morphology with a fixed latency and we use the retained
wavelet coefficients to provide both a detection statistic and a sparse
representation of the signal. By combining coefficients across consecutive analysis windows, the pipeline
reconstructs transients longer than a single window. We represent candidate
coincidences between detectors with a trigger graph and rank them with
supervised and unsupervised graph neural networks (GNNs). We estimate the
significance of the ranked coincidences from time shifted data. 
We validate the pipeline on Gaussian noise colored according to the Advanced LIGO sensitivity curve and on real O4 
strain containing compact binary and core collapse supernova injections. 
 Applied to GW250114, WDF identifies the event as the loudest zero lag candidate in the analyzed
segment without using waveform templates. We measure the computational cost of the pipeline and the
latency it introduces. Both are compatible with a low latency search, so WDF
can run on a stream in real time. 
\end{abstract} 
 
\keywords{gravitational waves, unmodeled transient search, wavelet transform,
whitening, low latency, graph neural network}

\maketitle
\section{Introduction}
\label{sec:introduction}

The observation of gravitational waves has opened a new way to study compact
objects and dynamical strong field gravity
\cite{abbott2016gw150914,abbott2017gw170817,gwtc3,gwtc4,gwtc5}. These
observations are made possible by the second generation interferometers
Advanced LIGO, Advanced Virgo and KAGRA
\cite{aligo2015,advirgo2015,kagra2021}. Most confident detections to date are
compact binary coalescences, for which accurate waveform models make matched
filtering the natural detection strategy. However, not all potential sources can be described by sufficiently
accurate waveform templates. Core collapse
supernovae, neutron star oscillations, post merger emission, cosmic string
bursts and unexpected short duration transients may have uncertain,
incompletely modeled or entirely unknown morphologies. Searches that do not
assume a predefined waveform family are therefore needed both to complement
matched filter analyses and to retain sensitivity to unexpected signals.

Unmodeled searches commonly identify excess energy or coherent signal
structure in a time frequency representation
\cite{anderson2001excesspower,klimenko2004wavelet,
chatterji2004qtransform,klimenko2016cwb,drago2021cwb,
lvk2021bursts,lvk2025bursts,martini2026cwb}. Production pipelines such as
coherent WaveBurst, BayesWave and Omicron use these representations to detect,
reconstruct or characterize transient candidates
\cite{klimenko2016cwb,bayeswave2015,robinet2020omicron}. Their sensitivity
does not depend on a detailed signal model. Without a template, they separate signals from instrumental transients using
two properties of a candidate: its consistency across the detector network and
its time frequency morphology.

The Wavelet Detection Filter (WDF) belongs to this class of searches. It assumes that, in a suitable wavelet basis, a short transient is sparse, or
at least more concentrated than the surrounding noise
\cite{mallat2009,daubechies1992}. WDF
selects significant wavelet coefficients in each analysis window and uses
their energy as a detection statistic. The same coefficients provide a sparse
representation from which a denoised waveform and physically interpretable
parameters can be reconstructed.

WDF was originally developed to search for short transients in Virgo data and
was later used as a front end for the automatic classification of instrumental
noise transients
\cite{cuoco2007grb,powell2015classification,
powell2017classification}. In the WDFX implementation
\cite{cuoco2018wdfx}, WDF derived features were combined with an XGBoost
classifier \cite{chen2016xgboost} to distinguish simulated glitch
morphologies from chirp like signals. The same classifier is used in coherent WaveBurst to rank burst candidates
from their time frequency features \cite{mishra2021xgboost,mishra2022xgboost}.
None of the earlier WDF versions estimated the accidental background,
combined the detectors of the network into a single significance, or
reconstructed a signal that extends over consecutive analysis windows. A WDF
trigger could therefore not be given a calibrated network significance.

The aim of this work is to develop WDF into a low latency pipeline for both
detection and waveform reconstruction. This requires three extensions: an
event built from the triggers of one detector, a calibrated significance for
that event and for the coincidences it forms, and a fixed latency for the
whole chain.

We analyze two data sets, a simulated two detector instrument and four
stretches of recorded O4 strain, to exercise the method and measure what it
recovers rather than to search for astrophysical signals.
GW250114 is analyzed as a check that the chain returns a known signal with no
template.

We first cluster the triggers of consecutive analysis windows into single
detector events, each carrying a \emph{wavegram}, a compact time frequency
representation of the wavelet coefficients its triggers retained. A signal longer than one analysis window is spread
over several such windows and its wavegram collects the coefficients from all of them. Inverting those
coefficients returns one waveform over the whole extent of the
signal, with each sample counted once. The events and the multi detector coincidences they form are then ranked
against an accidental background built by time slides. This background fixes
their significance and the detection efficiency of the search. The same
reconstructions also provide the timing of a coincidence. The arrival time difference measured from these reconstructions locates the
source on the sky, with a precision that the trigger times alone do not
provide. Throughout, the analysis is designed to operate with a
fixed and predictable latency as detector data arrive.

The interpretation of the WDF statistic depends directly on the conditioning
applied before the wavelet transform. WDF builds on the online autoregressive
noise identification and whitening framework originally introduced in
Ref.~\cite{cuoco2001online} and subsequently applied to interferometric
gravitational wave data in Ref.~\cite{cuoco2001ligo}. Here we introduce a
finite order, zero phase whitening filter that preserves transient morphology
and operates on a data stream with a fixed, predetermined latency. With this conditioning, the WDF statistic measures the norm of the
reconstructed waveform in units of the whitened noise and can therefore be
interpreted as the signal to noise ratio of the reconstruction.

The main methodological innovation of this work is the network ranking. The
events built in each detector form the nodes of a graph. Edges connect
events that satisfy the physical conditions required for a possible
coincidence, including compatibility in arrival time and in frequency
support. A connection between events in different detectors defines a candidate
gravitational wave coincidence. A conventional ranking reduces each candidate
to a few summary numbers, such as the loudness of the two events, their
arrival time difference and the consistency between the two detector records,
and combines them through a likelihood ratio whose form is fixed in advance
\cite{hanna2020gstlal}. We use a graph neural network (GNN) instead. It scores
a candidate from the descriptions of its two events, from what the edge
between them carries, and from the events neighbouring each of them in its own
detector \cite{gilmer2017mpnn}.

The GNN reads the sparse representation that WDF already produces. We train it
twice, once on injections on simulated data and once on accidental coincidences alone in real data. We threshold the resulting
score against the same time shifted background as the deterministic
statistic, so the rankings are compared at the same false alarm rate. To
our knowledge, this is the first use of a GNN to rank multi detector
coincidences directly on the trigger graph of an unmodeled transient search.

We validate the pipeline in two controlled settings: Gaussian noise colored
according to the Advanced LIGO sensitivity curve and selected stretches of
real O4 strain. Both contain compact binary and core collapse supernova injections. In
addition, the simulated set contains single detector transients of
instrumental morphology. These transients cannot be recovered
in coincidence and therefore measure the accidental floor of the network
stage.
The known injection times, classes and amplitudes allow the detection
efficiency, false alarm background and reconstruction accuracy to be measured
directly. We then apply the same pipeline to GW250114 as a
template independent test on a real gravitational wave signal. 

Section~\ref{sec:search-method} describes the workflow, conditioning, single
window statistic and construction of single detector events. The data sets, the analysis configuration
and the rule by which an injection counts as recovered are defined in
Sec.~\ref{sec:data}. Section~\ref{sec:network} constructs multi detector
coincidences and introduces the graph ranking.
Section~\ref{sec:network-results} reports the detection efficiency, for one
detector and for the network, on the simulated set and on real O4 strain,
compares the GNN and $R^{\mathrm{mor}}$ and gives the arrival time
accuracy and the sky localization that follows from it. Representative reconstructions are
presented in Sec.~\ref{sec:individual-signals} and discussed in
Sec.~\ref{sec:discussion}.  

\section{Search method}
\label{sec:search-method}

\subsection{Workflow and conditioning}
\label{sec:search}

The pipeline processes detector data as a continuous stream. The frames are
read sequentially, band passed and decimated to the analysis sampling
frequency. The stream is then whitened and analyzed in successive overlapping windows of
one fixed length. Each window is transformed, thresholded and scored
independently as it arrives. The pipeline does not require an entire data segment to be stored
before the analysis begins. The noise model is estimated before the segment is processed
and it is not recomputed inside the analysis loop for that segment.
Once the model is available, the conditioning, wavelet analysis and trigger production operate
sequentially on the incoming data.

\subsubsection{Zero phase whitening}
\label{sec:conditioning}

The conditioning must satisfy three requirements at once. First, the output
noise spectrum must be flat, so that the norm of the wavelet coefficients can
be expressed in units of a single noise scale. Second, the filter must not
distort the phase of a transient, since a distortion would affect both the
reconstructed waveform and the parameters derived from it. Finally, the filter
must run on a stream with a finite, predetermined latency, without
recomputing a spectral transform inside the processing loop.

The detector noise is modeled as white noise of scale $\sigma$ passed
through an all pole filter \cite{cuoco2001online,cuoco2001ligo}. For an
autoregressive model of order $p$,

\begin{equation}
    A(z) = 1-\sum_{k=1}^{p}a_k z^{-k},
    \qquad
    S_x(f) = \frac{\sigma^2}{\left|A(f)\right|^2},
    \label{eq:ar-noise-model}
\end{equation}

where $z^{-1}$ is the unit sample delay and $S_x$ is the noise power spectral
density. Whitening with $A$ flattens the spectrum but leaves a frequency
dependent phase and running a filter forward and then backward cancels that
phase at the price of applying it twice
\cite{kormylo1974zerophase,mattera2003noncausal,virtanen2020scipy}. The filter
applied in each direction is therefore the spectral square root of the fitted
model,

\begin{equation}
    \left|A_{1/2}(f)\right|^2 = \left|A(f)\right| ,
    \label{eq:square-root-filter}
\end{equation}

which we obtain as an autoregressive fit of order $q$ to the pseudo spectrum
$1/\left|A(f)\right|$. The conditioned stream is then white at zero phase,
with its noise scale recorded alongside it, and the backward pass needs
exactly $q$ future samples, so the look ahead is $q / f_s$, known before the
filter runs. Appendix~\ref{app:conditioning} gives the construction and the
conditioned spectrum in full.

\subsection{The single window WDF statistic}
\label{sec:single-window-statistic}

Each conditioned window is analyzed on its own. This subsection defines the
transform applied to it, the threshold that selects the coefficients to keep,
the statistic computed from those coefficients and the competition among
bases that decides which representation is used.

\subsubsection{The wavelet transform}

The search analyzes each window using ten orthonormal wavelet bases: the Haar
basis, centered Daubechies bases of orders 4, 8, 12, 16 and 20, Symlet bases of
orders 4 and 8, and Coiflet bases of orders 1 and 2
\cite{daubechies1992,mallat2009}. All transforms are implemented in the
\texttt{p4TSA} C++ core \cite{p4tsa}.

A window holds $N$ conditioned samples $\boldsymbol{x}$, and each basis $b$
is a complete orthonormal basis of that window: it contains $N$ waveforms
$\{\boldsymbol{\psi}^{(b)}_k\}_{k=0}^{N-1}$, obtained from its mother wavelet
at different dyadic scales and positions, together with the scaling function
of the lowest frequency band. The wavelet coefficients are

\begin{equation}
    w_k^{(b)}
    =
    \left\langle
    \boldsymbol{x},\boldsymbol{\psi}^{(b)}_k
    \right\rangle
    =
    \sum_{n=0}^{N-1}
    x_n\psi_{k,n}^{(b)},
    \qquad
    k=0,\ldots,N-1.
    \label{eq:wavelet-transform}
\end{equation}

The transform is repeated independently for all ten bases. Orthonormality
preserves the Euclidean norm and allows any selected set of coefficients to be
inverted exactly into the corresponding partial reconstruction.

\subsubsection{Thresholding and reconstruction}
\label{sec:thresholding}

For each basis, the scale of the block's own noise is estimated from the
median absolute deviation of its unthresholded wavelet coefficients,

\begin{equation}
  \sigma_b = \frac{\operatorname{median}_k \bigl( |w_k^{(b)}| \bigr)}{0.6745}.
  \label{eq:mad-noise}
\end{equation}

The coefficient threshold is the Donoho--Johnstone universal value
\cite{donoho1994},

\begin{equation}
    \lambda_b
    =
    \sigma_b\sqrt{2\ln N}.
    \label{eq:universal-threshold}
\end{equation}

WDF applies hard thresholding: a coefficient is kept at its own amplitude if it exceeds the threshold and is
set to zero otherwise,

\begin{equation}
    c_k^{(b)}
    =
    \begin{cases}
      w_k^{(b)}, & \left|w_k^{(b)}\right| \geq \lambda_b, \\[4pt]
      0,         & \text{otherwise.}
    \end{cases}
    \label{eq:coefficient-selection}
\end{equation}

The retained coefficients define the time domain reconstruction

\begin{equation}
    \hat{h}^{(b)}_n
    =
    \sum_{k=0}^{N-1}
    c_k^{(b)}\psi_{k,n}^{(b)}.
    \label{eq:wavelet-reconstruction}
\end{equation}

\subsubsection{The statistic}
\label{sec:statistic}

The WDF statistic for basis $b$ measures the total amplitude retained after
thresholding relative to the noise level in that basis. That level is read on
the blocks around the one being scored and never on the block itself: a block
containing a loud transient measures a scale the transient has raised and
dividing by it would hide the very thing the statistic is meant to see. The
retained coefficients are combined in quadrature,

\begin{equation}
    \EnWDF^{(b)}
    =
    \frac{
    \left\lVert\boldsymbol{c}^{(b)}\right\rVert_2
    }{\sigma},
    \label{eq:statistic}
\end{equation}

Since the transform is orthonormal, its inverse does not change this norm. If
$\hat{\boldsymbol{h}}^{(b)}$ is the waveform reconstructed from the retained
coefficients, Parseval's theorem gives

\begin{equation}
    \left\lVert\boldsymbol{c}^{(b)}\right\rVert_2
    =
    \left\lVert\hat{\boldsymbol{h}}^{(b)}\right\rVert_2.
\end{equation}

The same statistic can therefore be expressed in the time domain as

\begin{equation}
    \EnWDF^{(b)}
    =
    \frac{
    \left\lVert\hat{\boldsymbol{h}}^{(b)}\right\rVert_2
    }{\sigma}.
    \label{eq:snr}
\end{equation}

Thus, for correctly whitened data, $\EnWDF^{(b)}$ is an amplitude signal to
noise ratio, the amplitude of the reconstructed transient measured in units of
the noise. Hard thresholding leaves
the coefficients it keeps unscaled, so $\EnWDF^{(b)}$ equals the matched filter signal to noise ratio obtained when the
reconstruction is used as a template, an identity derived in
Appendix~\ref{app:matched-filter}.

A matched filter with a fixed template has a known distribution in noise.
This one does not, because the template is not fixed: the basis and the
coefficients that build it are chosen from the data being scored. We therefore
measure the distribution of $\EnWDF^{(b)}$ on the data instead of assuming it,
as described in Sec.~\ref{sec:background}.

\subsubsection{Basis competition}
\label{sec:bases}

Every window is analyzed in all ten bases, and each produces its own statistic.
The search keeps the largest of them,
\begin{equation}
    \EnWDF
    =
    \max_b
    \EnWDF^{(b)},
    \label{eq:basis-competition}
\end{equation}
and the corresponding basis provides the coefficients retained by the
trigger. The window produces a trigger if $\EnWDF$ exceeds the configured
trigger threshold. Basis competition allows each transient to be represented
in the basis that best concentrates its surviving coefficient norm.

The maximum of ten statistics computed on the same window is a different
random variable from any one of them: the pipeline is deterministic given the
data, but the data are not. In noise the maximum exceeds a fixed threshold more
often than a single basis does, since the ten statistics are only partly
correlated. No correction is applied for this. The background is measured with
the same ten basis search, so the null distribution it samples is that of the
maximum and statistics are compared at the same false alarm rate rather than at
the same numerical trigger threshold.

\subsection{Detector stage: building single detector events}
\label{sec:single-detector-events}

A transient need not fit inside one analysis window, so the triggers of a
single detector are grouped into events before any comparison between
detectors is made. This is the \emph{detector stage} and the cross detector
coincidence of Sec.~\ref{sec:network-stage} is the \emph{network stage}; the
two are named throughout by what they do and not by their order. This subsection builds those events from the windows,
defines the wavegram each of them carries, lists the parameters recorded with
a trigger and describes how the waveform is reconstructed across the windows
that an event spans.

\subsubsection{From analysis windows to events}
\label{sec:graphs}

The search reads the strain in consecutive analysis windows, or blocks, of a
single length. Each block shares a fixed stretch of overlapped samples with
the next. The runs reported here read blocks of $250$~ms overlapped by
$15.6$~ms, which is $512$ and $32$ samples at the $2048$~Hz analysis rate.
One analysis window length suffices because the dyadic transform is itself
multi-resolution:
with $N = 2^{J}$ samples, a coefficient at octave level $j$ covers the band
$[f_s 2^{\,j-J-1}, f_s 2^{\,j-J}]$ and lasts $2^{\,J-j}$ samples, so each band
is resolved on its own time scale. The overlap serves the reconstruction rather than the coverage of the data. A sample in the overlapped region has one
estimate from each of the two blocks that share it. Reconciling the two
estimates makes the crossing between blocks continuous once they are stitched
into one waveform.

A transient longer than one window, or one that falls astride a boundary,
leaves a trigger in each window it touches. We join those triggers into a
single event on geometry alone: two triggers belong to the same event when their energy lies close in time, when the octave bands they occupy touch and when
their energies are comparable. No criterion in the grouping favors a rising chirp or a compact burst.

The grouping is expressed as a graph. A node is a trigger, that is, the set of
coefficients retained in one window. For triggers $i$ and $j$,
let $d_t$ be the gap between the time supports of their retained energy,
$T_i$ and $T_j$ their window durations, $d_{\mathcal{B}}$ the separation of their
nearest occupied octave rows and $E=\EnWDF^2$ their retained coefficient
energy. An edge joins them when all three of

\begin{equation}
 \begin{aligned}
   &d_t\leq\tau_t\,(T_i+T_j)/2, \\[2pt]
   &d_{\mathcal{B}}\leq n_{\mathcal{B}}, \\[2pt]
   &\left|\log(E_i/E_j)\right|\leq\Delta_E
 \end{aligned}
 \label{eq:detector-edge}
\end{equation}

hold. The three tolerances are dimensionless: $\tau_t$ is the largest gap
allowed between two triggers as a fraction of their mean window duration,
$n_{\mathcal{B}}$ is the largest separation allowed in octave rows and $\Delta_E$
bounds the logarithm of the ratio of their energies. The analysis uses
$\tau_t=1$, so that two triggers may be separated by as much as one window
duration, $n_{\mathcal{B}}=1$, so that bands may touch without overlapping, and
$\Delta_E=3$, so that energies may differ by about a factor of twenty. We vary
each tolerance on its own by a factor of two in either direction and repeat the grouping from the same triggers. The recovered fraction at one false alarm per
day changes by at most one hundredth. The values above are not the result of an optimization over these tolerances. The
grouping does not
require the same wavelet basis in the two windows. All pairs that can satisfy
the time condition are tested and the connected components of the resulting
undirected graph are the single detector events. The construction is related
to the connected time frequency clustering of coherent WaveBurst
\cite{klimenko2016cwb} and to the wavelet graph representation of WaveGraph
\cite{bacon2018wavegraph,gayathri2019consistency}, but the graph here is built
from the observed WDF triggers and imposes no waveform morphology.

\subsubsection{The wavegram}
\label{sec:wavegram}

\begin{figure*}[!t]
  \centering
  \includegraphics[width=0.78\textwidth]{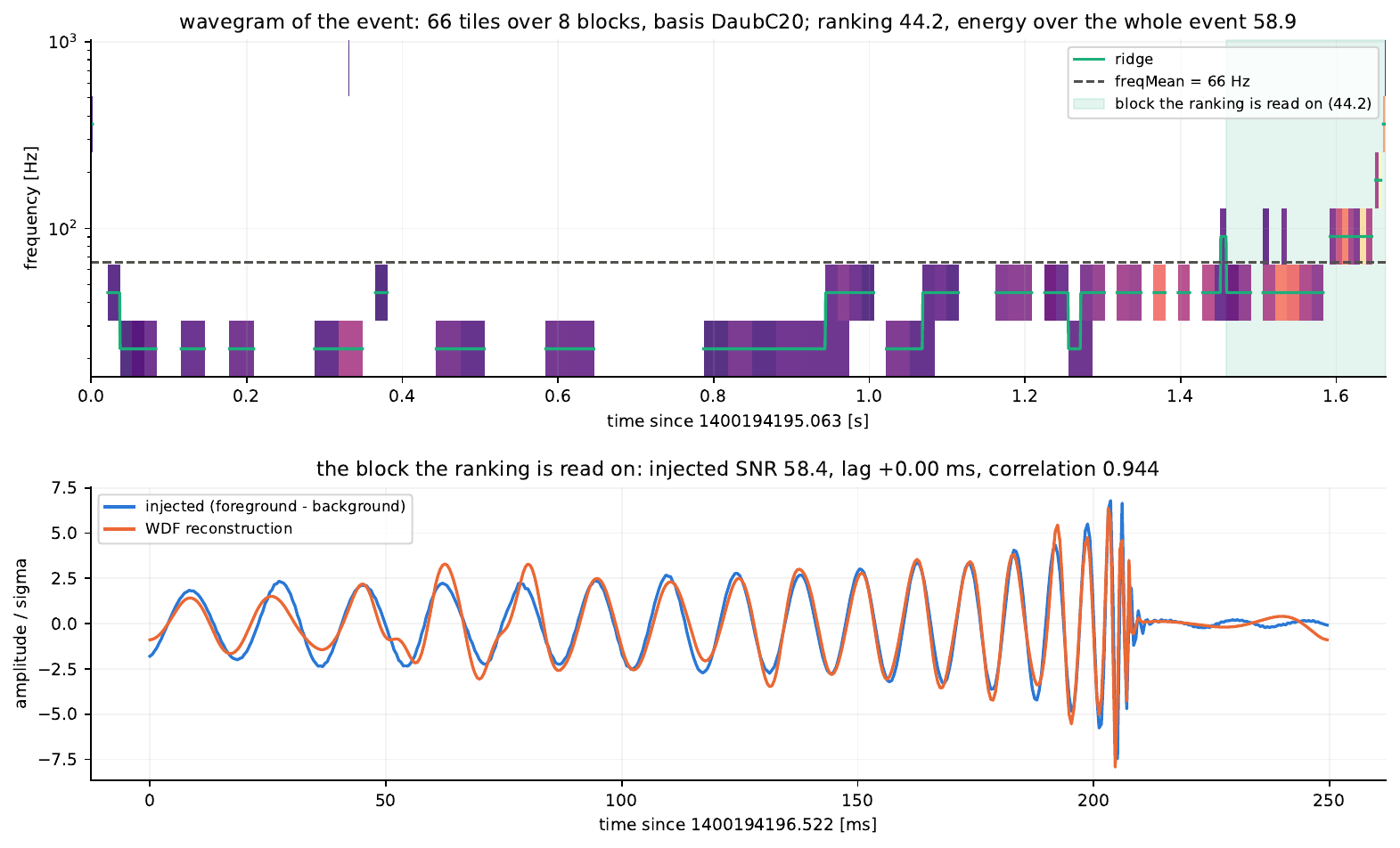}
  \caption{An injected binary black hole coalescence recovered in one
  detector, in its two representations. \emph{Top}: the event's wavegram, one
  rectangle per surviving coefficient over the band and time support its index
  implies, shaded by $\lvert c_k \rvert / \sigma$, with the ridge solid and
  the mean frequency dashed. The event spans eight analysis windows; the shaded
  stretch is the block the detection statistic is read on; the two numbers in
  the title are that block's value and the norm over the whole event.
  \emph{Bottom}: within that block, the injected waveform, taken as foreground
  minus background, against the reconstruction inverted from the same
  coefficients, with the correlation between them.}
  \label{fig:wavegram}
\end{figure*}

A coefficient index fixes a time interval and an octave band. For $N=2^J$, a
coefficient $k\geq1$ belongs to octave $j=\lfloor\log_2 k\rfloor$: its band is
$[f_s2^{j-J-1},f_s2^{j-J}]$ and its time support is $2^{J-j}$ samples, while
$k=0$ is the scaling coefficient. The \emph{wavegram} of an event is the
collection of its retained coefficients rendered as tiles on a common
absolute time frequency plane, read off the orthonormal transform itself.
Because the transform is orthonormal and the data are whitened,
$\lvert c_k \rvert / \sigma$ is the signal to noise ratio carried by tile $k$. We take the loudest tile in each time bin and these tiles form the
ridge of the wavegram, which is the time frequency track of the transient
\cite{carmona1997ridge}. Figure~\ref{fig:wavegram} shows the construction. Related
wavelet graph representations preserve transient structure across scales
\cite{chassandemottin2017waveletgraphs}.

The rows of a wavegram are indexed by absolute frequency band, so that the
same physical band is the same row in every detector. Each event therefore
carries a description of its own time frequency morphology, of the kind previously used to separate gravitational wave transients from
instrumental ones
\cite{powell2015classification,powell2017classification,cuoco2018wdfx}.

A trigger is a pair: the coefficients that survived thresholding, with the
indices that place them, and the noise scale $\sigma$ measured on that window.
Everything downstream is a function of that pair. The time a transient is
centered on, its extent and its band are moments of the energy its tiles carry,
and the wavegram, the coincidence tests and the reconstruction read the same
coefficients. None of them returns to the strain.
Appendix~\ref{app:parameters} lists the parameters recorded with each trigger.

\subsubsection{Reconstruction across windows}
\label{sec:multiwindow}

A sample in the overlapped region has an estimate from each of the two windows
that cover it. The two estimates are not equal, because each window
thresholded its own coefficients. Reconciling them is the analysis--synthesis problem of a
modified short time transform. Its least squares solution is a weighted
overlap-add \cite{allen1977unified,crochiere1980wola,griffin1984stft}. We write
$\hat x_w$ for the inverse transform of window $w$ and $g$ for a weight over
its samples, and obtain
\begin{equation}
  \hat{h}_{\mathrm{stitched}}(t) =
  \frac{\sum_w g(t - t_w)\, \hat x_w(t - t_w)}
       {\sum_w g(t - t_w)} ,
  \label{eq:overlapadd}
\end{equation}
which counts each sample once, however many windows cover it.

No window is applied at analysis, so $g$ is a free choice of the stitching. It
must vanish at the block edges: the two estimates of an overlapped sample
disagree after thresholding, so a weight that does not vanish leaves a step at
every boundary. We use a raised cosine over the overlapped samples, flat in
between, which is the smallest such choice, and whose two ramps are
complementary, so that the interior is not rescaled. The exact least squares
weight \cite{griffin1984stft} reduces here to the unweighted mean, which is
discontinuous wherever the set of covering windows changes. We choose
continuity over exact optimality.

The result is a single time series covering the whole signal. Because the
noise is white, \EnWDF\ can be read on this series exactly as it is read on
one window,
\begin{equation}
  \EnWDF = \frac{\lVert \hat{h}_{\mathrm{stitched}} \rVert_2}
                {\sigma},
  \label{eq:stitched}
\end{equation}
with $\sigma$ the noise scale of the blocks the event spans.
This is the same statistic as Eq.~\eqref{eq:snr}, measured over the signal's
full extent rather than over one window.

The two steps act in sequence. The grouping of the detector stage selects the
windows that belong to the event, and the stitched reconstruction of
Eq.~\eqref{eq:overlapadd} gives its amplitude. The stitching weights sum to one
on every sample, so a sample covered by two windows is counted once. Summing the
energies of the tiles instead would count a tile once per window that covers
it. The value reported for the event is therefore $\EnWDF$ of
Eq.~\eqref{eq:stitched}, and its time and frequency span is the extent of the
tiles belonging to the event.

The stitched waveform never carries the whole of the signal, whatever weight is
used: only the windows that produced a trigger contribute, so the stretches of
an inspiral that stay below threshold add nothing. This is a property of a
thresholded search and not of the stitching. Recovering them requires knowing
in advance where the signal lies, which a coherent or a template driven
follow-up supplies and an unmodeled search does not.

Figure~\ref{fig:wavegram} shows the two representations of one event side by
side, its wavegram and the waveform its coefficients invert to, against the
signal that was injected.

Everything described so far involves a single detector. Used in this way, the search
is already a noise characterization tool. It triggers on transients of every
morphology without a model of any of them and each event carries the wavegram
and the reconstruction that a classification chain takes as input.

\section{Data sets and recovery criterion}
\label{sec:data}

This section describes the data sets on which the pipeline is evaluated: a
simulated two detector set with its noise and its injections, a second
realization of it that is held out and the stretches of real O4 strain on
which the analysis is repeated.

\subsection{Simulated}
\label{sec:mock}

The simulated set describes two detectors of comparable sensitivity. Their
noise is colored Gaussian at the \texttt{aLIGOZeroDetHighPower} design curve,
and each frame kind carries $2.98$~days of livetime. The set is generated at
twice the analysis rate.

We inject three kinds of signal and measure how the search performs on each.
Compact binary coalescences are the sources observed so far. Short
instrumental transients are the most common contaminant of a burst search.
Core collapse supernovae are an astrophysical signal with no closed form. The morphologies are chosen to
span the time frequency plane rather than to reproduce any one catalog. In
total we place $6590$ injections and we record for each one its time, class,
parameters and optimal signal to noise ratio:

\begingroup
\squeezetable
\begin{table}[!htbp]
  \centering
  \caption{The five single detector noise transient morphologies of the
  simulated set. Time $t$ runs from the center of the waveform, and
  $\tau = Q/2\pi f_0$ is the decay time of the envelope. Every parameter is drawn
  uniformly over the range listed, independently for each injection. The
  amplitude is set separately, by scaling the waveform to a single detector
  signal to noise ratio drawn between $8$ and $100$. For the chirp like and
  scattered light classes the entry gives the instantaneous frequency and the
  envelope. The waveform is the envelope times the sine of the accumulated
  phase.}
  \label{tab:glitches}
  \begin{ruledtabular}
    \begin{tabular}{lll}
      class & waveform & parameters \\
      \colrule
      Gaussian & $e^{-t^{2}/2\sigma_t^{2}}$
               & $\sigma_t$: $2$--$20$~ms \\[2pt]
      sine-Gaussian & $e^{-t^{2}/2\tau^{2}}\sin(2\pi f_0 t)$
                    & $f_0$: $60$--$600$~Hz \\
                    & & $Q$: $5$--$30$ \\[2pt]
      blip & as above, rise $\tau$, decay $3\tau$
           & $f_0$: $80$--$500$~Hz \\
           & & $Q$: $2$--$5$ \\[2pt]
      chirp like & sweep $f_1 \to f_2$ linear in $t$,
                 & $f_1$: $25$--$80$~Hz \\
                 & Gaussian envelope & $f_2$: $150$--$700$~Hz \\
                 & & duration: $0.2$--$1.5$~s \\[2pt]
      scattered light & $f(t) = f_{\mathrm{p}}\lvert\sin(\pi t/T)\rvert$
                      & $f_{\mathrm{p}}$: $20$--$60$~Hz \\
                      & & $T$: $0.5$--$2$~s \\
                      & & arches: $2$--$5$ \\
    \end{tabular}
  \end{ruledtabular}
\end{table}
\endgroup

\begin{itemize}
  \item \textbf{compact binaries}, generated with
        \textsc{IMR\-Phenom\-D}~\cite{husa2016,khan2016} for the binary black
        holes and \textsc{Taylor\-F2}~\cite{taylorf2} for the systems
        containing a neutron star ($3349$ binary black holes, $258$ black
        hole--neutron star binaries and $193$ binary neutron stars), each
        projected onto each detector through its own antenna response and
        light travel delay. Sky position, inclination and polarization are
        drawn isotropically and the network signal to noise ratio between $8$
        and $100$. How it divides between the detectors follows from the
        source's position and we require that neither detector be left below
        a signal to noise ratio of $7$. The amplitude ratios and arrival time differences the coincidence stage is
        tested against are those of that population rather than the full range
        the geometry allows;
  \item \textbf{2600 noise transients}, or glitches, each in a
        \emph{single} detector, so any coincidence involving a glitch is a false alarm by construction. They
        measure the accidental floor of the network stage and one detector's
        recovery is read on them as on any other injection, over morphologies
        no compact binary or supernova provides. We use five closed form
        morphologies, chosen to span the plane. Table~\ref{tab:glitches} defines
        them: a broadband pulse with no carrier frequency, the narrowband
        sine-Gaussian, the short and broadband blip that instrumental
        transients resemble, a rising frequency sweep, and scattered light,
        which is a train of low frequency arches lasting seconds. The
        arches and the longer sweeps outlast the analysis window, so the detector stage must assemble 
        them from several of its blocks;
  \item \textbf{190 core collapse supernovae}, drawn from the catalog of
        three dimensional, initially non-rotating simulations of
        \cite{choi2024ccsn} and computed as described in
        \cite{vartanyan2023ccsn}. This is the class on which an unmodeled search is compared with the targeted
supernova searches
        \cite{szczepanczyk2024ccsn}. A supernova waveform has no closed form and no parameters to draw. We draw both the progenitor model and the observer direction relative to the
simulation's axes uniformly over the catalog. The two polarizations are those produced by the simulation, projected onto
each detector through its
        antenna response and light travel delay. The catalog's own time steps are resampled to the
        analysis rate with a band limited filter, so that emission above the
        analysis Nyquist frequency is removed rather than folded into the
        band. Each waveform is also high passed at $10$~Hz and
        brought to zero at its end before it is injected. Several recent three dimensional campaigns
        provide waveforms of this kind
        \cite{powell2019ccsn,powell2020ccsn,vartanyan2023ccsn}. We do not compare
        these campaigns. We use a single catalog in order to place in the data a signal that has no
closed form and against which the reconstruction
        can be read.
\end{itemize}

The set is generated twice, with two seeds and one configuration. The first
realization is where the rankings are fitted and the tolerances chosen; the
second is held out and the network results of Sec.~\ref{sec:network-sim} are
read on it. It carries its own $6590$ injections, drawn from the same
population.

Each realization is written as two frame sets on one noise realization. Both
carry the instrument's transients, at the same times and in the same
detectors, and the \emph{foreground} set adds the astrophysical signals. The
difference of the two is therefore exactly the injected waveform, and every
trigger is attributed to a signal or to the noise without recourse to time
slides.

\subsection{Real O4 strain}
\label{sec:realdata}

Real strain is neither Gaussian nor stationary and it carries instrumental
transients that the simulated set does not reproduce. We therefore repeat the analysis on real O4 strain.
We use four stretches from the O4b open data release \cite{gwosc_o4b}.
Each stretch is analysis ready in both LIGO detectors, and the stretches are
separated by days, so that no two share a noise realization. They are the stretches of April 8, April 13, April 19 and April 21--22, 2024,
starting at GPS $1396581517$,
$1397001846$, $1397543013$ and $1397768900$ and spanning between $11.8$ and
$18.2$ hours each. The first falls in the engineering segments that the
release includes, before the official start of O4b on April 10, 2024. The other three lie within the run itself. Its transient catalog is GWTC-5.0
\cite{gwtc5}.

The injections are the astrophysical population of the simulated set: compact
binaries generated with \textsc{IMRPhenomD} and \textsc{TaylorF2} and core
collapse waveforms drawn from the catalog, placed isotropically on the sky
and projected onto each detector through its antenna response and light
travel delay, with the network signal to noise ratio measured against each
stretch's own spectrum. 

We add the signal alone to the strain and the noise is the detector's own,
untouched. Each stretch is written twice, \emph{foreground} with the
injections and \emph{background} without, exactly as for the simulated set, so
the two frame sets differ by the injected waveforms and nothing else. No two
injections overlap: each reserves its own support and a clear gap around it.

Together the four stretches hold $2.16$~days of analyzed livetime per frame kind and $3915$ astrophysical
injections: $3281$ binary black holes, $263$ black hole--neutron star
systems, $184$ binary neutron star systems and $187$ core collapse
supernovae. We inject no noise transients. The real strain carries its own transients
and they enter the background as the detectors produced them.
The AR noise model is fitted per stretch and applied to that stretch
alone.

\subsection{Configuration and matching rule}
\label{sec:configuration}

The search is configured the same way on both data sets; what differs is only
that the autoregressive noise model is fitted once per simulated segment and
once per recorded stretch. The strain is decimated by two to an analysis rate
of $2048$~Hz and band-passed above $12$~Hz with a Chebyshev type II filter of
order $10$. The noise model is autoregressive of order $p = 3000$, fitted on
$300$~s of strain and it is applied through the zero phase square root filter
of Eq.~\eqref{eq:square-root-filter} at order $q = 256$; four blocks are
conditioned before the search starts, so that the filters have settled on
everything the search reads. The transform runs on the blocks of
Sec.~\ref{sec:graphs} and a window produces a trigger when \EnWDF\ exceeds
$5$. An injection and a candidate are matched within $50$~ms.

An injection counts as recovered when its time falls inside the stretch a
candidate covers, widened by the matching window; among the candidates that
cover it, the one recorded is the loudest on the statistic being read.
Matching is on the
candidate's extent and not on its own instant: an event's time sits where its
energy is and a chirp carries most of its energy before the merger, so a test
on instants would miss a long signal that the search recovered correctly. Each
injection takes at most one candidate and the injections are placed so that
no two overlap, so no candidate stands for more than one injection. Every
fraction reported below is the number of injections recovered under this rule,
divided by the number injected.

\section{Network coincidence and graph ranking}
\label{sec:network}

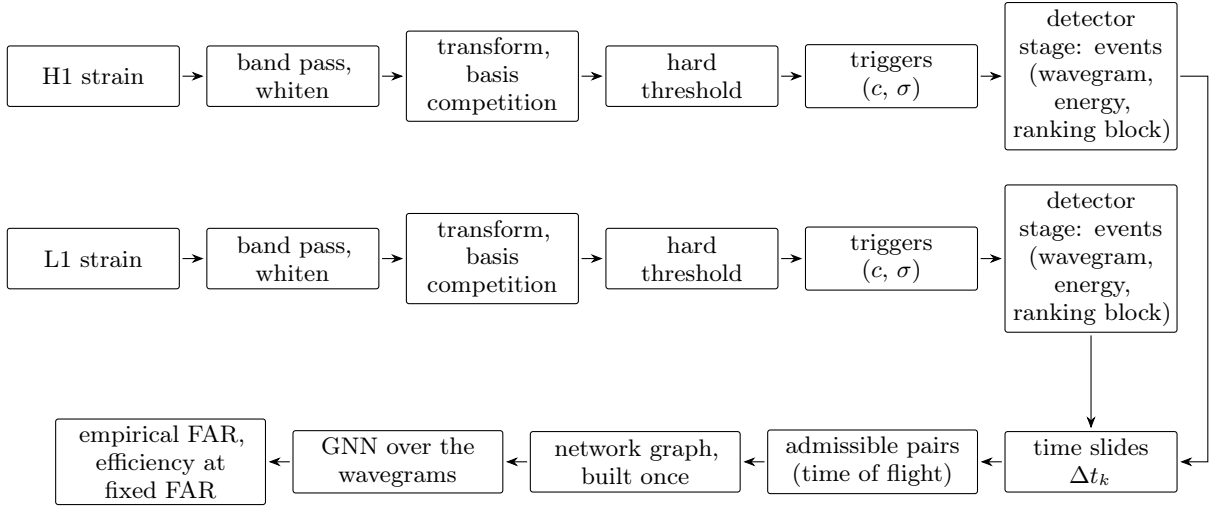
\begin{figure*}[!t]
  \centering
  \begin{tikzpicture}[
    font=\small,
    box/.style={draw, rounded corners=1pt, align=center, inner sep=2.5pt,
                minimum height=8mm, text width=21mm},
    wide/.style={box, text width=26mm},
    >={Stealth[length=1.6mm]},
    line/.style={->, shorten >=1pt, shorten <=1pt},
  ]
    \matrix[column sep=3.5mm, row sep=4.5mm] {
      \node[box] (h1) {H1 strain}; &
      \node[box] (c1) {band pass,\\whiten}; &
      \node[box] (t1) {transform, basis\\competition}; &
      \node[box] (s1) {hard\\threshold}; &
      \node[box] (g1) {triggers\\$(c,\,\sigma)$}; &
      \node[box] (e1) {detector stage: events\\(wavegram, energy,\\ranking block)}; \\
      \node[box] (h2) {L1 strain}; &
      \node[box] (c2) {band pass,\\whiten}; &
      \node[box] (t2) {transform, basis\\competition}; &
      \node[box] (s2) {hard\\threshold}; &
      \node[box] (g2) {triggers\\$(c,\,\sigma)$}; &
      \node[box] (e2) {detector stage: events\\(wavegram, energy,\\ranking block)}; \\
    };
    \node[box, below=13mm of e2] (slide)
      {time slides\\$\Delta t_k$};
    \node[wide, left=3.5mm of slide] (pairs)
      {admissible pairs\\(time of flight)};
    \node[wide, left=3.5mm of pairs] (graph)
      {network graph,\\built once};
    \node[wide, left=3.5mm of graph] (rank)
      {GNN over the\\wavegrams};
    \node[wide, left=3.5mm of rank] (far)
      {empirical FAR,\\efficiency at fixed FAR};
    \foreach \a/\b in {h1/c1, c1/t1, t1/s1, s1/g1, g1/e1,
                       h2/c2, c2/t2, t2/s2, s2/g2, g2/e2,
                       slide/pairs, pairs/graph, graph/rank, rank/far}
      \draw[line] (\a) -- (\b);
    \draw[line] (e1.east) -- ++(4mm,0) |- (slide.east);
    \draw[line] (e2.south) -- (slide.north);
    \node[font=\bfseries, anchor=south west]
      at ([yshift=2.5mm]current bounding box.north west) {The WDF pipeline};
  \end{tikzpicture}
  \caption{Each detector's strain is conditioned, transformed at one
  window length under the basis competition and hard thresholded. The
  triggers are grouped into events by the detector stage and for each shift
  $\Delta t_k$ the pairs that the time of flight admits become the cross
  detector edges of a graph the GNN ranks. The grouping precedes the shifts,
  so the events do not change with them and zero lag is the same construction
  at $\Delta t = 0$.}
  \label{fig:pipeline}
\end{figure*}

The same construction runs independently in every detector of the network and
produces that detector's own events. The grouping of Sec.~\ref{sec:graphs} is
a graph inside one detector, whose nodes are triggers; the network stage builds
a second graph, across detectors, whose nodes are the single detector events
themselves. It admits an edge only where a coincidence is physically possible.
The admitted pairs are then ranked and a pair's rank becomes a false alarm
rate by counting how many accidental pairs of the time slides reach it.
Figure~\ref{fig:pipeline} summarizes the whole chain, from each detector's
strain to the rate.

\subsection{Network stage: multi detector coincidences}
\label{sec:network-stage}
\label{sec:networkresults}

\begin{figure}[!tb]
  \centering
  \includegraphics[width=\columnwidth]{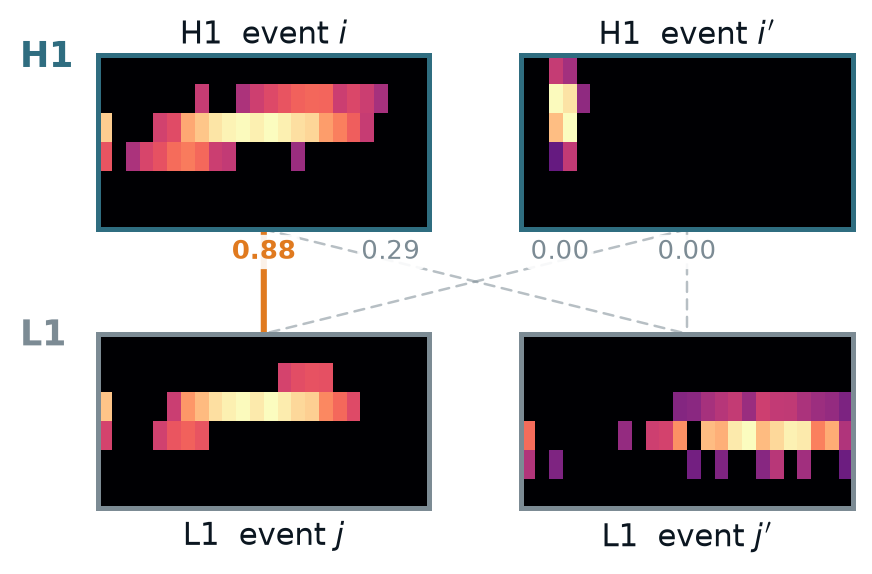}
  \caption{The cross detector part of a trigger graph, on a constructed example
  with two events in each detector. A node is one detector stage event, shown
  as its wavegram, with time running left to right and frequency band upward.
  An edge joins an event $i$ of one detector to an event $j$
  of the other and is a candidate coincidence, carrying the arrival
  time difference and the agreement between the two wavegrams; the number
  drawn on an edge is that agreement and the solid edge is the pair of
  largest agreement. Each wavegram is normalized before it is compared, so the
  agreement is between shapes and not between amplitudes, which the antenna
  responses make unequal between detectors.}
  \label{fig:trigger-graph}
\end{figure}

An edge joins two events in different detectors only where a signal could
have produced the pair, and Fig.~\ref{fig:trigger-graph} shows the resulting
graph. The two must cover the same stretch of time once one of them is shifted
by a tolerance $\Delta_{ij}$: the light travel time between the sites, widened
by the events' own timing spreads and held to a small multiple of it. That
upper limit keeps $\Delta_{ij}$ a property of the network geometry: the arrival
times of one signal differ by at most that time, while a timing spread is an
uncertainty on an event's own centroid that is largely common to the two
detectors. An unbounded tolerance would admit a pair of long events over
seconds, so the accidental rate would follow the duration of the events
instead of the geometry. Overlap in band is not imposed: the shared fraction of band is a
feature of the edge that the ranking reads, so a pair is never rejected on its
bands. These are the candidates a conventional coincidence test
\cite{usman2016pycbc} admits, so the GNN and the deterministic statistics rank
one population and can be compared.

The test is on the events' extents rather than on an instant: for a transient
shorter than the light travel time the two coincide, so the extent test is the
general one and the arrival time difference is left to rank the survivors
rather than to decide which pairs exist.

The pipeline uses two times that are not interchangeable. The first is the
event's own \emph{instant}, the peak of the analytic envelope of its stitched
reconstruction; the edge carries the difference of two such instants. That
time belongs to one event, so a time slide moves it with the event and the
accidental population costs no measurement per pair. The second is the lag that maximizes the
cross-correlation of the two reconstructions. It belongs to the pair and is
measured only on the candidates that survive and it is used for the sky localization.

An edge is described by the following quantities: the arrival time
difference in seconds and as a fraction of the tolerance $\Delta_{ij}$; the
shared fraction of band and the shared fraction of time support; the log ratio
of the two energies; the agreement between the two wavegrams and the overlap
of the two grids; the deterministic coincidence statistics themselves. 
A shared fraction is the overlap of the two intervals divided by the shorter
of their widths, so an interval contained in the other shares all of it.

Let $W$ be the wavegram of Sec.~\ref{sec:wavegram} read as a vector, one
entry per cell and zero where no coefficient survived. Its entries span
decades, so we take the logarithm of each of them before comparing two
wavegrams. On that scale a cell counts more for its presence than for its
loudness, and an empty cell still sits at exactly zero, so the grids stay as
sparse as the thresholding made them. The agreement between two events $i$
and $j$ is then the cosine of the angle between the two transformed vectors,
\begin{equation}
  s_{ij} = \frac{\bigl\langle \log(1 + W_i) , \, \log(1 + W_j) \bigr\rangle}
                {\lVert \log(1 + W_i) \rVert_2 \,
                 \lVert \log(1 + W_j) \rVert_2} ,
  \label{eq:shape}
\end{equation}
which is $1$ when the two grids have the same shape, whatever their
amplitudes.
Because each grid has its time axis referred to that event's own energy
centroid, the two are aligned by construction and
Equation~\eqref{eq:shape} compares shape alone, leaving the arrival time
difference (the difference of the two events' envelope instants) and the
energy ratio to travel on the edge as features of their own, so that geometry
and amplitude enter the ranking once, explicitly.
Nothing bounds the energy ratio, while the arrival time difference is bounded
by the light travel time between the sites, which is the tolerance
$\Delta_{ij}$ the pair was admitted on.

A candidate's false alarm rate is measured against the accidental
coincidences that the time slides produce \cite{was2010timeslides}. The same
statistic is applied to the accidentals and to the zero lag candidates. The livetime of
those slides sets the lowest rate that can be quoted, so a candidate louder
than every accidental is quoted at that rate as a limit. Efficiency is measured against a floor rather than as a raw fraction.
Injections placed in one detector only cannot be recovered in coincidence, so
the fraction of them that a statistic appears to recover is produced by
chance matching alone.

\subsection{Deterministic ranking}
\label{sec:deterministic}

The first of the two rankings is deterministic: a fixed formula, with no
training behind it, that orders the admitted pairs by the agreement of their
morphologies and not by loudness alone. We define it on the coefficients
themselves and use it as the reference the GNN is measured
against. For a pair made of event $i$ in one detector and event $j$ in the
other,
\begin{equation}
  R^{\mathrm{mor}}_{ij} = \Biggl\lvert \sum_{(k,l)}
      \frac{c^{(i)}_k}{\sigma_i} \, \frac{c^{(j)}_l}{\sigma_j}
      \Biggr\rvert ,
  \label{eq:rank-morphology}
\end{equation}
where $c^{(i)}_k$ is the $k$-th surviving coefficient of event $i$ and
$\sigma_i$ the noise scale it was measured on, so that $c^{(i)}_k / \sigma_i$
is the signal to noise ratio that tile carries, and $(k,l)$ runs over the pairs
of tiles of the two events whose bands overlap and whose time supports meet
within the light travel time.
$R^{\mathrm{mor}}$ is therefore a coherent energy, in units of the noise
squared. It requires not only that both
detectors were loud, but that they were loud in the same places on the plane,
which is the comparison the wavegram representation makes possible.

\subsection{Graph neural network ranking}
\label{sec:gnn}
\label{sec:learned}

The network stage produces a graph and not a list of pairs. A graph neural
network is a model that learns on a graph directly, reading the features of
its nodes and of its edges together with the way they are connected
\cite{scarselli2009gnn,battaglia2018relational}. We therefore train one on that graph and use it to
rank the candidates in place of the deterministic statistic of the network
stage, whose functional form is fixed in advance. Nothing upstream changes. The conditioning, the thresholding, the detector
stage grouping and the admissibility rule are those of the preceding
sections; only the ordering of the admitted pairs is learned. The graph is the one
already shown in Fig.~\ref{fig:trigger-graph}. A node is an event of one detector, a panel of that figure: it carries the
coefficients that the event kept, rendered on a shared band-by-time grid as
its wavegram, together with the scalars that the search measured.
A cross detector edge is a pair that the admissibility rule admitted. In the figure it is drawn between the
panels and labeled by the agreement of Eq.~\eqref{eq:shape}. An event is
joined, inside its own detector, to the events the grouping of
Sec.~\ref{sec:graphs} placed with it, so that its neighborhood in the graph is
the transient it belongs to. These intra-detector edges are what let a node be
read together with its surroundings rather than alone.

Graph neural networks have entered gravitational wave
analysis on the strain, over a fixed graph whose nodes are the detectors, for
modeled binary signals \cite{tian2024gnn}. Learned searches read the strain itself and answer a different question: a
coincidence and coherence test across detectors \cite{mly2024}, or anomaly
detection against a learned model of the background \cite{gwak2024}. Those
approaches learn a representation of the data, whereas this ranking learns a
relation between triggers that the search has already produced. For the broader landscape of
machine learning in this field see \cite{cuoco2025mlreview}.

\subsubsection{Message passing}
\label{sec:mpnn}

Through \emph{message passing} \cite{gilmer2017mpnn} a single model reads
graphs of any size, whose number of nodes and pattern of connections change
from one input to the next. Each node starts from its own features and, in every
round, updates its state from the states of its neighbors and of the edges
joining them, so that, after a few rounds, its embedding describes the node in
its context rather than in isolation.
The same learned functions are applied at every node and every edge, so the
model does not depend on how many events a stretch of data produced or on how
they are connected. This suits a trigger graph, whose size and connectivity
change from segment to segment while its local structure does not: an event
among its neighbors in its own detector, compared with an event in the other
detector.

On the trigger graph, every node is embedded and messages travel along the
intra-detector edges. The two embeddings of a candidate pair are then combined
with the features that the edge carries and the model returns a single number
for the pair. 

\subsubsection{Supervised and unsupervised training}
\label{sec:gnn-training}

Two configurations are used. They differ in the data they are trained on,
not in architecture. The \emph{supervised} configuration is trained on
injections in simulated data sets. A cross detector edge is labeled one when a compact binary falls within its
matching window and zero otherwise. The positives are therefore coincidences
that a source produced and the negatives are everything else that the
admissibility rule admitted. The model learns from them what a source
produces across two detectors. It is fitted on the simulated set and applied
unchanged to real O4 strain, a background it never saw. Training on real
strain needs a labeled set built in the instrument's own noise and is
reported separately \cite{cuoco2026inprep}.

The \emph{unsupervised} configuration is an anomaly detector: it is trained on
accidentals alone and sees no signal at all. Time slides, or equivalently
frames containing no injection, give a population that is accidental by
construction and can be made as large as the background requires. An
autoencoder is fitted on it. For each candidate edge it compresses the two
events' embeddings and, from the compressed form alone, reproduces the
features $\bm{x}$ that the edge carries. Having seen accidentals only, it
does this well for pairs that resemble them and badly for pairs that do not,
so the size of the reconstruction error measures how anomalous a pair is
against the accidental population and that error is the score. A candidate is
therefore ranked by how unlike an accidental coincidence it is, and never by
how like a signal it is.

A likelihood ratio would compare the density of $\bm{x}$ under a signal with
its density under the accidental hypothesis. The search has no signal model,
so only the second of the two can be built and the reconstruction error
stands in for the whole ratio. The price is that the autoencoder learns the
noise of the stretch it was fitted on and does not carry to another one, so
the background that fits it must not be the background that calibrates it.
Neither output is a probability of astrophysical origin. Both are ranking
statistics, calibrated and compared exactly as the deterministic ones are.

\section{Detection performance}
\label{sec:network-results}
\label{sec:performance}

Efficiency is reported twice, on the simulated set and on a stretch of real O4
strain, for single detector and for the network of two. Each rate is read off the
background of the stage that produced it and each efficiency is quoted with the
livetime that resolves it. Time slides~\cite{was2010timeslides} extend
the livetime over the same noise.

\subsection{Background}
\label{sec:background}

The background frames carry the same noise realization as the foreground and
differ from it by the injections alone, so the events they produce are a draw
from the null distribution of the statistic. Rates are read off that draw and
never extrapolated: a threshold is quoted as a rate, never as the maximum of a
finite stretch, whose loudest event is a single draw from the tail. The
livetime therefore fixes the lowest rate a single detector can state, one
background event in the stretch analyzed.

\subsection{Results on the simulated set}
\label{sec:sim-performance}

\subsubsection{Single detector}

On the simulated set the search triggers on blips, scattered light,
sine-Gaussians, chirp like transients and broadband Gaussian bursts, with no
model of any of them. The basis selected by the competition differs between
these classes. The search was run over the $2.98$~days of the set. A candidate
is ranked on $\EnWDF$ measured on its loudest block, and we report alongside
it the event's energy, the norm over the whole event measured on the stitched
reconstruction.

With no threshold attached, the search recovers $0.99$ of the binary black
holes in each detector, $0.87$ and $0.86$ of the black hole--neutron star
systems, $0.47$ and $0.50$ of the binary neutron stars and $0.60$ and $0.63$ of
the core collapse waveforms: $0.94$ of the astrophysical population in either
detector.
The instrumental morphologies enter one detector at a time, so a coincident
search is not asked to recover them; the same statistic recovers them at the
level of the astrophysical classes, between $0.85$ for scattered light and
unity for the blips, the Gaussians and the sine-Gaussians.

The ordering of the classes follows from thresholding block by block. A binary
black hole deposits most of its energy in the few blocks around the merger, so
it is almost always recovered. A binary neutron star sweeps over far longer
than one block: little of it survives threshold in any single one, which is
why it is recovered least.

The single detector results show that the search responds to morphologies whose durations
differ by three orders of magnitude, with no model of any of them, and that
the coefficients it keeps invert back to the waveform.

\subsubsection{The network}
\label{sec:network-sim}

The network stage is read on a \emph{held-out} realization of the simulated
set: a second, independent $2.98$~days drawn with a different seed under the
same mixture, which no fit has seen, so the reading is out of sample for the
GNN and on equal terms for $R^{\mathrm{mor}}$. Five hundred
time slides of the same events turn those $2.98$~days into $1489.6$~days of
accidental livetime. The lowest rate this background resolves is one accidental
event in that livetime, five hundred times lower than the unslid $2.98$~days
would allow.

\begin{figure*}[!t]
  \centering
  \includegraphics[width=\textwidth]{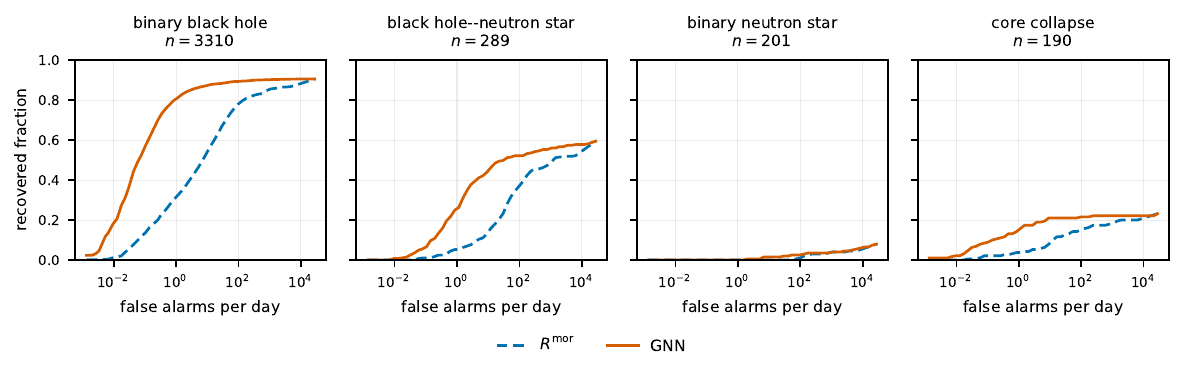}
  \caption{Fraction of the coincident injected population recovered against
  false alarm rate, one panel per class, on the held-out simulated set. The
  dashed curves show the deterministic ranking $R^{\mathrm{mor}}$ and the solid
  curves the supervised GNN. Both are read on the same candidates and against
  the same background.}
  \label{fig:mock-network-efficiency}
\end{figure*}

Table~\ref{tab:mock-network-class} and Fig.~\ref{fig:mock-network-efficiency}
report the recovered fraction on the held-out simulated set, class by class and
rate by rate, for the deterministic ranking $R^{\mathrm{mor}}$ and for the
GNN. The GNN leads at every rate the background resolves and the
lead is largest where recovery is hardest: on binary black holes it recovers
$0.81$ against $0.31$ at one false alarm per day, and $0.40$ against $0.06$
at one per month. The two draw closer at the permissive end without meeting,
$0.78$ against $0.89$ on binary black holes at one hundred per day, and on the
classes the analysis window treats worst neither ranking finds much: on binary
neutron stars both stay at a few hundredths at every rate.

\begin{table*}[!t]
  \centering
  \caption{Fraction of the coincident injected population recovered on the
  held-out simulated set, by class and by false alarm rate, for the
  deterministic ranking $R^{\mathrm{mor}}$ of Eq.~\eqref{eq:rank-morphology} and for the supervised GNN.
  Both are read on the same admitted candidates and against the same
  background of $1489.6$~days of accumulated time-slide livetime. $n$ is the
  number of injections in each class.}
  \label{tab:mock-network-class}
  \begin{ruledtabular}
    \begin{tabular}{lrrrrrrrrr}
      class & $n$ & \multicolumn{2}{c}{$1$/month} & \multicolumn{2}{c}{$1$/day} & \multicolumn{2}{c}{$10$/day} & \multicolumn{2}{c}{$100$/day} \\
      \cline{3-10}
       &  & $R^{\mathrm{mor}}$ & GNN & $R^{\mathrm{mor}}$ & GNN & $R^{\mathrm{mor}}$ & GNN & $R^{\mathrm{mor}}$ & GNN \\
      \colrule
      binary black hole          &  3310 & 0.06 & 0.40 & 0.31 & 0.81 & 0.54 & 0.87 & 0.78 & 0.89 \\
      black hole--neutron star   &   289 & 0.00 & 0.03 & 0.05 & 0.26 & 0.14 & 0.44 & 0.37 & 0.52 \\
      binary neutron star        &   201 & 0.00 & 0.00 & 0.00 & 0.00 & 0.00 & 0.01 & 0.01 & 0.02 \\
      core collapse              &   190 & 0.01 & 0.07 & 0.04 & 0.15 & 0.09 & 0.21 & 0.15 & 0.22 \\
      all coincident             &  3990 & 0.05 & 0.34 & 0.27 & 0.70 & 0.46 & 0.77 & 0.68 & 0.79 \\
    \end{tabular}
  \end{ruledtabular}
\end{table*}

\subsection{Results on real O4 strain}
\label{sec:real-performance}

\subsubsection{Single detector}
\label{sec:realdata-single}

The same search runs unchanged on the four stretches of real O4 strain. What
a single detector can claim on real noise follows from the background of those
stretches.

The loudest background events of real O4 strain are instrumental transients
whose loudest block carries $\EnWDF$ in the hundreds, so a threshold read at any rate this livetime resolves sits far
above every injection: it characterizes the tail of the glitch distribution
and not the search
\cite{davis2021detchar,powell2015classification,powell2017classification}.
What one detector does establish is how much of what was injected it recovers.
With no threshold attached, over the $2.16$~days of recorded strain, the
search recovers $0.995$ of the binary black holes in H1 and $0.999$ in L1,
$0.943$ and $0.966$ of the black hole--neutron star systems, $0.799$ and
$0.859$ of the binary neutron stars, and $0.524$ and $0.572$ of the core
collapse waveforms; over the whole injected population, $0.960$ in H1 and
$0.970$ in L1. On real O4 strain a single detector therefore serves as a
trigger generator, and no detection is claimed from it alone.

\subsubsection{The network}
\label{sec:o4injections}

The network stage runs on the same four stretches. Time slides of the
recorded events, up to five hundred displacements of each stretch, give
$1082$~days of accidental livetime. Of the $3915$ injections, $352$
have no admitted pair. Of those, $146$ have an event in both detectors and
were refused by admissibility, and $206$ are missing an event in at least one
detector. The admitted fraction is the ceiling of every efficiency reported
below, and it is not one number: $0.99$ of the binary black holes, $0.86$ of
the black hole--neutron star systems, $0.33$ of the binary neutron stars and
$0.18$ of the core collapse waveforms. A class recovered at a few hundredths
is read against its own ceiling and not against the $0.91$ of the population
as a whole.

\begin{figure*}[!t]
  \centering
  \includegraphics[width=0.82\textwidth]{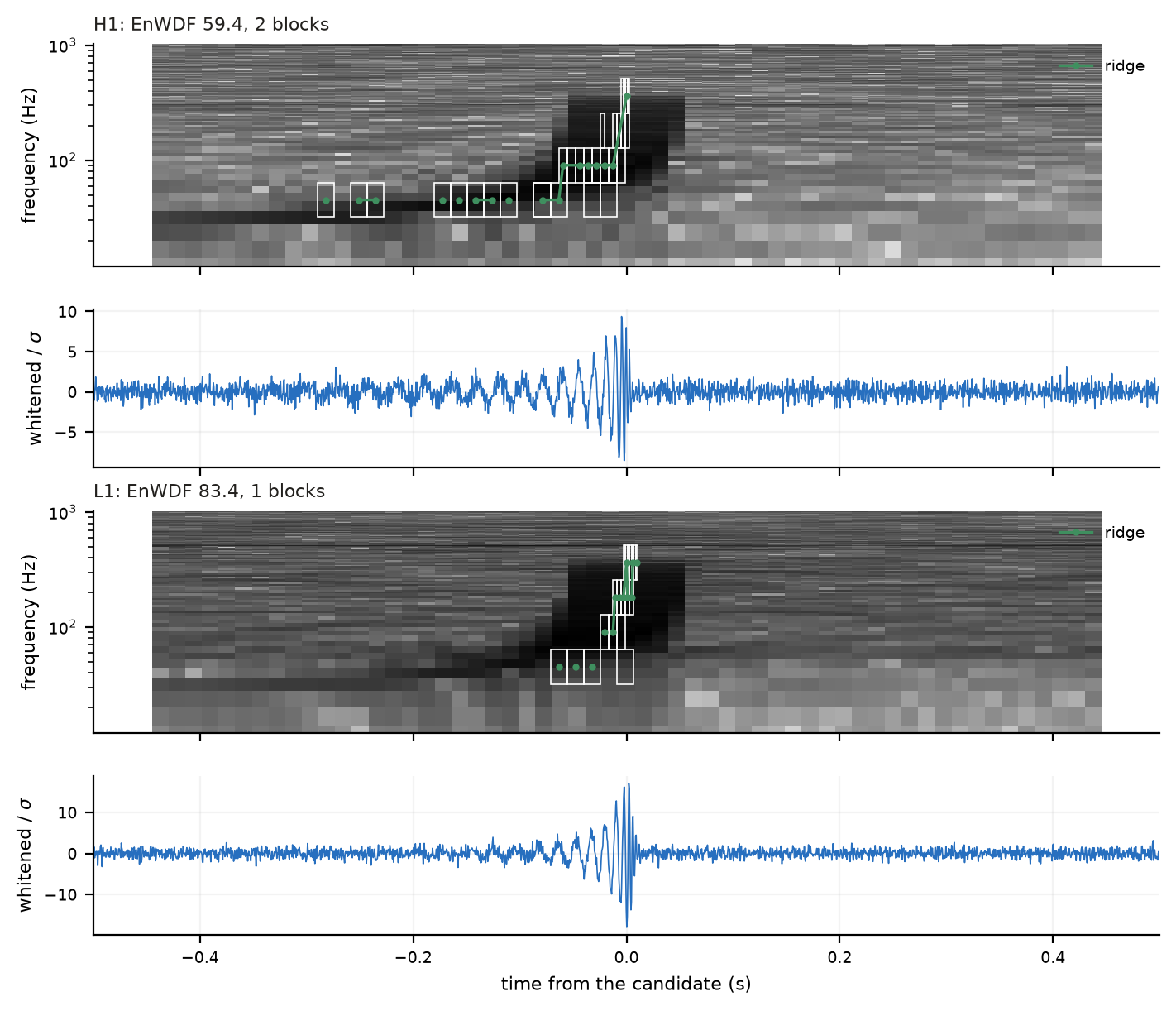}
  \caption{An injected binary black hole recovered in coincidence on real O4
  strain, H1 above and L1 below. For each detector, the upper panel shows the
  time frequency plane around the candidate, with the tiles that the search
  kept outlined and the ridge on which their energy lies. The lower panel
  shows the whitened strain that contains the candidate. The arrival time
  difference of the pair is $-2.9$~ms, inside the light travel time. The event
  is assembled from two blocks in H1 and one in L1, so in H1 the ridge runs
  through every tile those blocks kept, the low frequency ones a quarter of a
  second from the merger included: the detector stage joins blocks whose energy
  is close in time without requiring their bands to meet.}
  \label{fig:recovered-pair}
\end{figure*}

The accidental livetime, and not the analyzed span, fixes the rates that the
network can quote. One false alarm per month is resolved by the background
itself, with no extrapolation and no threshold set on a maximum, but at that rate
no injection of the $3915$ survives under either ranking, so the reading starts
at one false alarm per day. Every class is read at the network's own rate, since an
accidental pair carries no class. Figure~\ref{fig:recovered-pair} shows one
admitted candidate as the search represents it: an injected binary black hole
recovered in both detectors, with the arrival time difference within the light
travel time.

\begin{figure*}[!t]
  \centering
  \includegraphics[width=\textwidth]{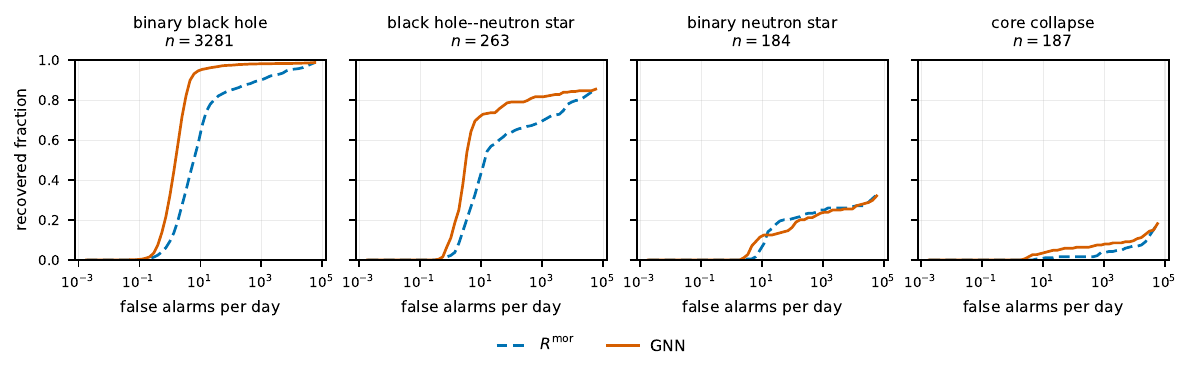}
  \caption{Fraction of the coincident injected population recovered against
  false alarm rate, one panel per class, on real O4 strain. The dashed curves
  show the deterministic ranking $R^{\mathrm{mor}}$ and the solid curves the
  supervised GNN. The model is the one fitted on the simulated set and is not
  refitted here.}
  \label{fig:real-network-efficiency}
\end{figure*}

Table~\ref{tab:real-network-class} and Fig.~\ref{fig:real-network-efficiency}
repeat the reading on recorded strain, with the model fitted on the simulated
set and not refitted. The ordering is the same, with the GNN above
$R^{\mathrm{mor}}$: $0.95$ against $0.63$ on binary black holes at ten false
alarms per day, and $0.98$ against $0.85$ at one hundred per day. That livetime resolves one false alarm per month, and at that rate the
recovered fraction is zero in every class, so the table starts at one per
day.

\begin{table*}[!t]
  \centering
  \caption{Fraction of the coincident injected population recovered on real
  O4 strain, by class and by false alarm rate, for the same two rankings as
  Table~\ref{tab:mock-network-class}. At one false alarm per month, which this
  background resolves, no injection of the $3915$ is recovered, so the table
  starts at one per day. The last column is the loosest rate reported, where
  the efficiency saturates and the two rankings agree. $n$ is the number of
  injections in each class.}
  \label{tab:real-network-class}
  \begin{ruledtabular}
    \begin{tabular}{lrrrrrrrrr}
      class & $n$ & \multicolumn{2}{c}{$1$/day} & \multicolumn{2}{c}{$10$/day} & \multicolumn{2}{c}{$100$/day} & \multicolumn{2}{c}{$1000$/day} \\
      \cline{3-10}
       &  & $R^{\mathrm{mor}}$ & GNN & $R^{\mathrm{mor}}$ & GNN & $R^{\mathrm{mor}}$ & GNN & $R^{\mathrm{mor}}$ & GNN \\
      \colrule
      binary black hole          &  3281 & 0.09 & 0.32 & 0.63 & 0.95 & 0.85 & 0.98 & 0.90 & 0.98 \\
      black hole--neutron star   &   263 & 0.02 & 0.11 & 0.44 & 0.72 & 0.64 & 0.79 & 0.70 & 0.82 \\
      binary neutron star        &   184 & 0.00 & 0.00 & 0.05 & 0.11 & 0.21 & 0.17 & 0.25 & 0.23 \\
      core collapse              &   187 & 0.00 & 0.00 & 0.01 & 0.03 & 0.02 & 0.06 & 0.04 & 0.07 \\
      all coincident             &  3915 & 0.08 & 0.28 & 0.56 & 0.85 & 0.77 & 0.88 & 0.82 & 0.89 \\
    \end{tabular}
  \end{ruledtabular}
\end{table*}

\subsection{Supervised and unsupervised GNN}
\label{sec:statistic-selection}

The GNN is fitted in two ways, on labeled injections and on the
background alone. The supervised model is the one reported above, fitted on
the simulated set's injections against its accidental pairs. The unsupervised
model is a graph autoencoder trained on the first part of every stretch and
thresholded on the second; the two do not overlap, so its reading is a
measurement and not a bound.

\begin{figure*}[!t]
  \centering
  \includegraphics[width=0.49\textwidth]{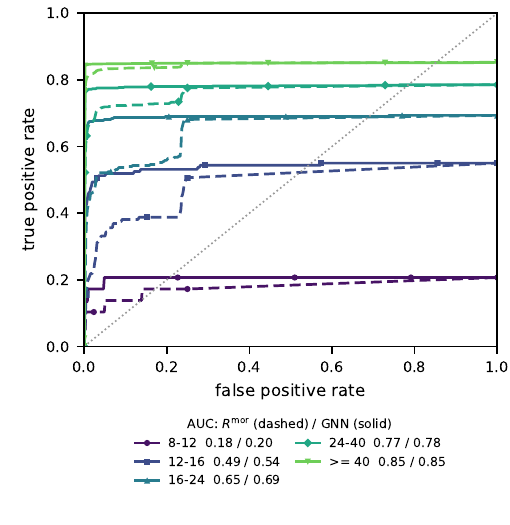}\hfill
  \includegraphics[width=0.49\textwidth]{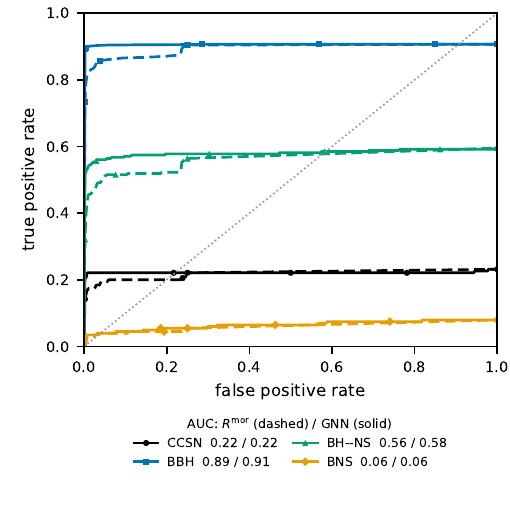}
  \caption{Receiver operating characteristic of the admitted coincidences on
  the held-out simulated set, one curve per band of injected network signal to
  noise ratio (\emph{left}) and per class (\emph{right}). The deterministic
  statistic is dashed and the supervised GNN solid, in the colour of the band
  or class they share and the legend gives the two areas under the curves in
  that order, the classes abbreviated as BBH, BH--NS, BNS and CCSN. Both axes are fractions of a population, one trial being one
  accidental pair; an injection matched by no candidate is a miss at every
  threshold, so a curve saturates at the fraction of its class the two
  detectors admitted in coincidence. The populations are those of
  Table~\ref{tab:mock-network-class}. The fit saw only the first set, so the
  reading is out of sample.}
  \label{fig:mock-network-roc}
\end{figure*}

\begin{figure*}
  \centering
  \includegraphics[width=0.49\textwidth]{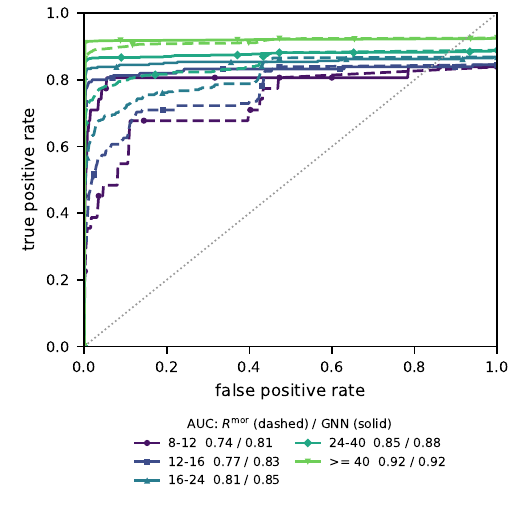}\hfill
  \includegraphics[width=0.49\textwidth]{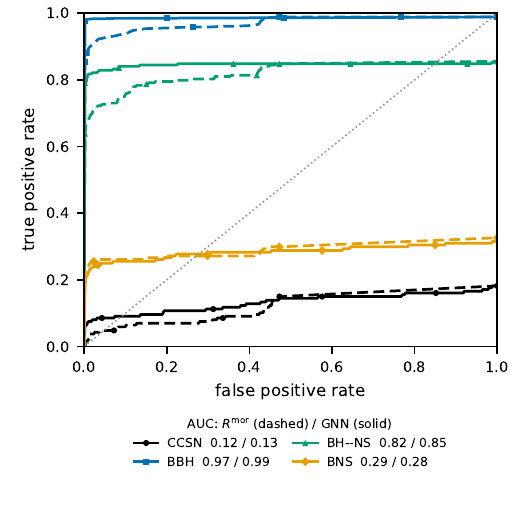}
  \caption{As Fig.~\ref{fig:mock-network-roc}, on real O4 strain. The GNN is
  trained on the simulated set only and applied here unchanged.}
  \label{fig:network-roc}
\end{figure*}

Read on the slid background of real O4 strain, the supervised ranking
recovers $0.28$ of the coincident population at one false alarm per day and
$0.85$ at ten per day, against $0.08$ and $0.56$ for $R^{\mathrm{mor}}$.
The unsupervised score is read on a background of its own, the unslid second
parts. At a false alarm probability of $10^{-2}$ the three are within a tenth
of one another --- $0.89$ for the supervised ranking, $0.82$ for the
unsupervised and $0.80$ for $R^{\mathrm{mor}}$, against a ceiling of $0.91$ --
and at $10^{-3}$ the unsupervised one collapses to $0.04$ while the other two
hold, $0.88$ and $0.75$. Learning from labeled injections is worth twenty
points of efficiency at one false alarm per day; learning from the background
alone holds its own where the probability is loose and falls away where it is
tight, because the accidentals above its threshold there are pairs of loud
glitches.
The three are not read on backgrounds of the same depth. The autoencoder is
calibrated on $51\,484$ accidental pairs of the unslid second parts, which
resolve a probability of $1.9\times10^{-5}$, while the other two are read on
the slid population, whose livetime resolves one false alarm per month. The
unsupervised ranking is therefore quoted at a probability on its own
background and not at a rate beside the other two. We adopt the supervised ranking, with the deterministic statistic
beside it as the reference a hand written formula sets.
Figures~\ref{fig:mock-network-roc} and~\ref{fig:network-roc} show the two
rankings as receiver operating characteristics over the unit square, per band
of injected network signal to noise ratio and per class, on the held-out
simulated set and on real O4 strain.

\subsection{Arrival time and sky localization}
\label{sec:skyloc}

The arrival time difference underlies both the coincidence and the sky
position. We measure it against the injection times recorded for each detector
separately. The arrival time difference of a candidate is measured on its two
reconstructions. The two events are inverted and stitched, and the two
reconstructions are placed on one absolute time grid. The arrival time
difference is the lag that maximizes their cross-correlation, searched within
$\pm 250$~ms, and its uncertainty is the half width of that peak above half
its maximum, floored at one sample. The pair is timed on the waveform that the
two detectors share, at the sample.

The estimator costs little. The stitched series of an event is already
computed for the whole event statistic of Eq.~\eqref{eq:stitched}, so the
timing adds one regridding and one cross-correlation per candidate.

Table~\ref{tab:timing} reports the residual of the arrival time difference in
bands of injected network signal to noise ratio. On the simulated set the
spread is $0.59$~ms above network signal to noise ratio $40$ and $0.98$ of
the residuals fall inside the light travel time; on real O4 strain the spread
reaches $0.44$~ms above $40$ and $0.98$ of the residuals fall inside the
light travel time. The tail of the residuals comes from candidates
whose two detectors kept different parts of the transient. The
cross-correlation peaks sharply only where the two reconstructions share their
morphology and glitches share none. A signal of effective bandwidth $\sigma_f$ received at matched filter
signal to noise ratio $\rho$ can in principle be timed to
$\sigma_{\Delta t} = 1/(2\pi\rho\sigma_f)$
\cite{fairhurst2009triangulation,fairhurst2011localization}, which is
$40$~$\mu$s at that loudness and at a bandwidth of a hundred hertz, so the
error we measure stands an order of magnitude above what the signal allows.

\begin{table}[!htbp]
  \centering
  \caption{The error on the arrival time difference, measured on the two
  stitched reconstructions, in bands of injected network signal to noise ratio.
  ``Spread'' holds the central 68 per cent of the errors and ``in time'' is the
  fraction of them smaller than the light travel time.}
  \label{tab:timing}
  \begin{ruledtabular}
    \begin{tabular}{lrrr}
      band & $n$ & spread & in time \\
      \colrule
      \multicolumn{4}{l}{\emph{simulated set}} \\
      8--12     &   12 & 5.6~ms & 0.83 \\
      12--16    &   75 & 2.5~ms & 0.80 \\
      16--24    &  232 & 0.85~ms & 0.93 \\
      24--40    &  549 & 0.66~ms & 0.99 \\
      $\geq 40$ & 2390 & 0.59~ms & 0.99 \\
      \colrule
      \multicolumn{4}{l}{\emph{on real O4 strain}} \\
      8--12     &   26 & 2.1~ms & 0.88 \\
      12--16    &  131 & 0.61~ms & 0.98 \\
      16--24    &  291 & 0.50~ms & 0.97 \\
      24--40    &  615 & 0.47~ms & 0.98 \\
      $\geq 40$ & 2500 & 0.44~ms & 0.99 \\
    \end{tabular}
  \end{ruledtabular}
\end{table}

\begin{figure*}[!t]
  \centering
  \includegraphics[width=\textwidth]{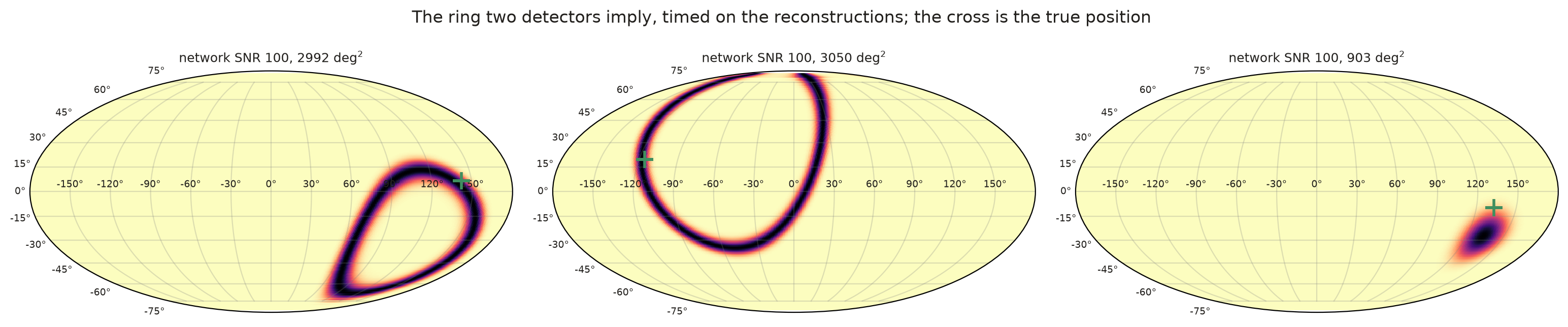}
  \caption{The sky region implied by the three loudest coincident candidates
  recovered on real O4 strain, all at network signal to noise ratio close to
  $100$, timed on the two stitched reconstructions. Each ring's width is the
  spread the residuals of that band of loudness show in Table~\ref{tab:timing},
  $0.44$~ms here, and the cross marks the injected position. The three rings
  cover $2992$, $3050$ and $903$ square degrees.}
  \label{fig:real-sky}
\end{figure*}

The sky region is built from that spread. A difference measured to $\sigma$
places the source on a ring whose half width is
$\arcsin(\sigma/\Delta t_{\max})$ where the source is perpendicular to the
baseline and wider elsewhere, so the $0.59$~ms of the simulated set is a ring
of $3.4$ degrees at best, and $3.8$ and $4.9$ degrees in the two bands below.
The uncertainty given to each pair is the spread that its own band of loudness
shows in Table~\ref{tab:timing}, measured on the same data set as the pair.
Over the $3244$ recovered binary black holes of the real O4 strain, the true
position falls inside the $90$ per cent region for $2821$ of them, $0.87$ of
the population, and between $0.86$ and $0.92$ of the time in every band,
against the $0.90$ claimed. The median area is $3.0\times 10^{3}$ square
degrees against a sky of $4.1\times 10^{4}$.
Figure~\ref{fig:real-sky} shows the regions of the three loudest recovered
candidates, each about $3.4\times 10^{3}$ square degrees wide.

\section{Signal reconstructions}
\label{sec:reconstruction-results}
\label{sec:individual-signals}

\subsection{A core collapse supernova}
\label{sec:ccsn-reconstruction}

\begin{figure}[!ht]
  \centering
  \includegraphics[width=\columnwidth]{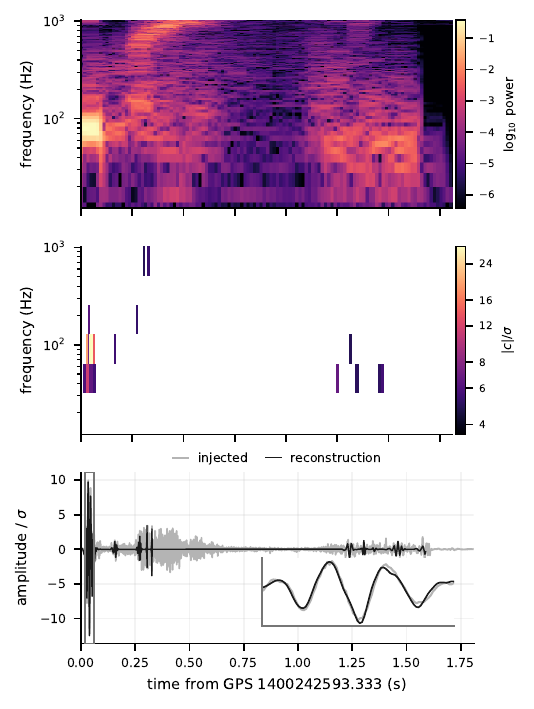}
  \caption{The core collapse supernova of the simulated set, injected and
  recovered. From the top: the injected signal on the time frequency plane,
  conditioned as the search input is; the wavegram of the recovered event, the
  coefficients that survived thresholding in its four analysis windows; and the
  injected waveform against the stitched reconstruction, which agree with a
  correlation coefficient of $0.97$. Every panel begins at the bounce, and the
  inset magnifies the $40$~ms marked on the last one. The strain files are the
  public ones of \cite{choi2024ccsn}, at \url{https://dvartany.github.io/data/}.}
  \label{fig:ccsn-waveform}
\end{figure}

The stitched reconstruction is not specific to
the chirps that make up most of the injected population.
Figure~\ref{fig:ccsn-waveform} applies it to a core collapse waveform of the
simulated set, a morphology that no part of the search was built around. The
figure shows the injected signal on the time frequency plane, the wavegram of
the recovered event and the injected waveform against the reconstruction
assembled from the analysis windows that the event spans. The progenitor is the $40\,M_\odot$ model of
the three dimensional simulations of \cite{choi2024ccsn}, computed as described
in \cite{vartanyan2023ccsn} and seen along the simulation's $y$ axis. It is
injected at network signal to noise ratio $98.2$, and the recovered energy
concentrates where the signal crosses the sensitive band.

\subsection{GW250114}
\label{sec:gw250114}

We read $34$~minutes of public strain from LIGO Hanford and LIGO Livingston
around the reported time of GW250114 \cite{gw250114}, released through the open
data program \cite{gwosc_o3,gwosc_o4,gwosc_o4b}, of which $23$~minutes are
analyzed once the conditioning has settled. The search is the one of
Sec.~\ref{sec:search} in the configuration of Sec.~\ref{sec:configuration},
with the autoregressive model refitted to this noise. The segment is too short
to support a false-alarm rate worth quoting, and no significance statement
about the event is made or implied here.

\begin{figure*}[!t]
  \centering
  \includegraphics[width=\textwidth]{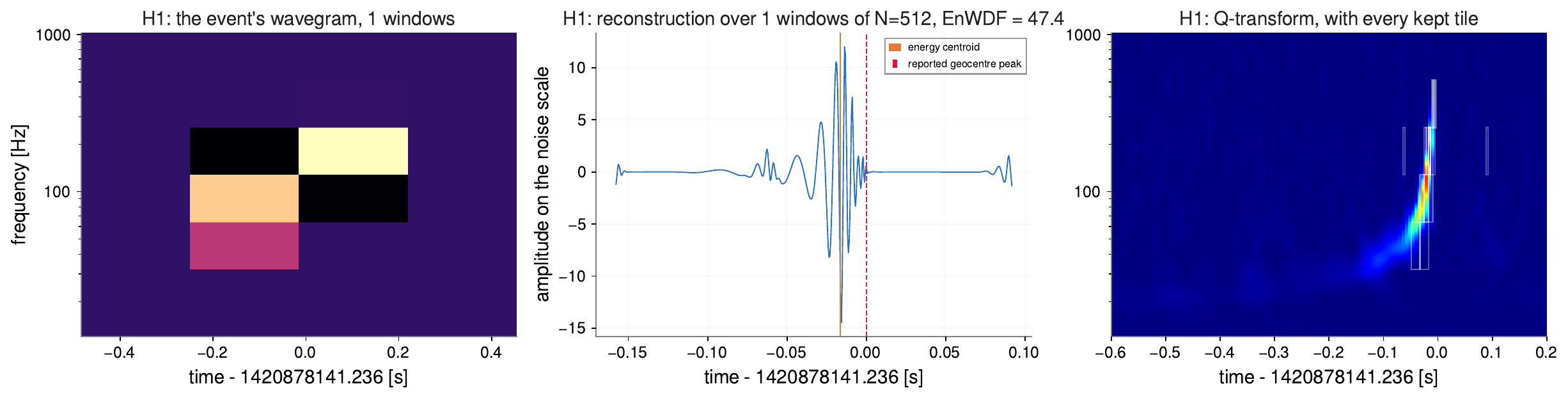}\par\medskip
  \includegraphics[width=\textwidth]{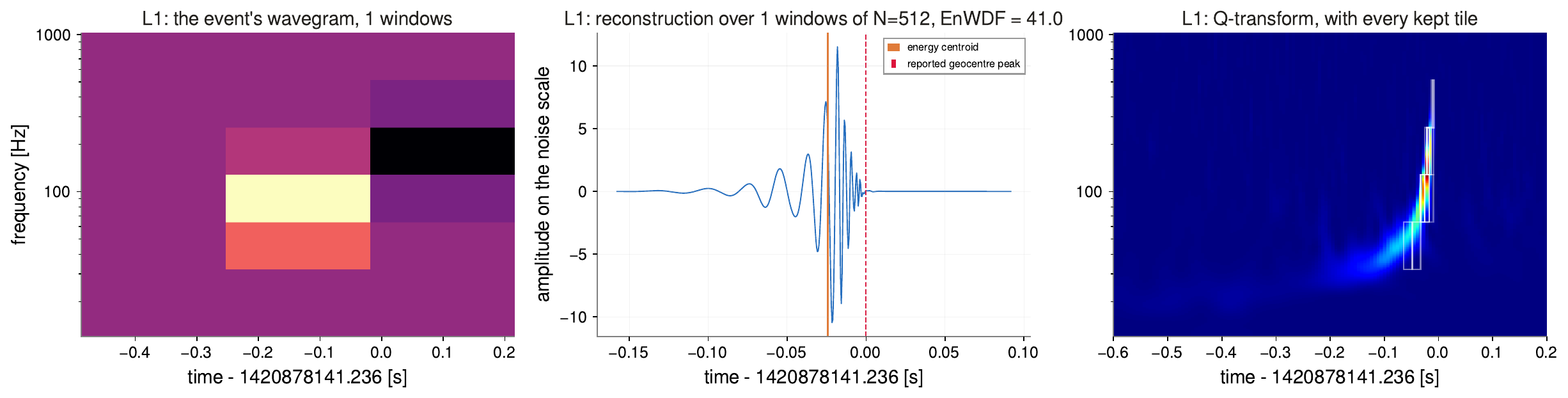}
  \caption{The event covering GW250114 in each detector, H1 above and L1
  below, in the basis that won the competition on that block, DaubC16 in H1
  and DaubC20 in L1. \emph{Left}: the coefficients that survived thresholding,
  fifteen in H1 and twelve in L1, each drawn over the band and the time
  support its index implies. \emph{Center}: the waveform they invert to, on
  the noise scale, with the event's energy centroid and the reported peak at
  the geocenter marked. \emph{Right}: a $Q$-transform of the same strain,
  computed independently, with every kept tile drawn over it.}
  \label{fig:gw250114}
\end{figure*}

The detector stage grouped the triggers of each detector into events by the
geometric admissibility rules, with no reference to the reported time.
Figure~\ref{fig:gw250114} shows the resulting event in each detector. Both
energy centroids precede the reported peak at the geocenter, GPS
$1420878141.236$ \cite{gwosc}, by $17$~ms in H1 and $24$~ms in L1, of the
order of the light travel time to it, and they lie within the light travel
time between the two sites. A chirp
deposits its energy before the merger, so a centroid ahead of it is expected.

The figure shows what the search kept. The analysis window spans a quarter of
a second at this rate and the transient fits inside one window. The detector
stage therefore assembled an event of a single window in each detector and
the statistic over the event's extent equals the statistic of that window.
The grouping does not recover this signal and is not required to. A binary
black hole at this mass is shorter than one block and the stage that matters
here is the coincidence.

At the network stage the two detectors' events formed a candidate pair, with
their arrival times $2.93$~ms apart against a light travel time of $10.01$~ms.
Ranked on $R^{\mathrm{mor}}$, it is the loudest zero lag candidate of the
segment.

GW250114 is reported with a network matched filter signal to noise ratio of
about 80 and a false alarm rate far below what this segment resolves
\cite{gwosc}. \EnWDF\ measures excess energy, a different quantity from the
match to a template and each ranking is read on the background of its own
search. The background here is extended by time slides. One thousand
displacements are asked for, each at least $10$~s and applied to the span the
search covered; a stretch this short admits $123$ distinct ones, which give
$1.79$~days of accidental coincidences, a livetime that resolves $0.56$ per
day. The candidate stands above every accidental pair of that background, so
its rate is quoted as a limit, below $0.56$ per day, set by the livetime this
segment supports.

\FloatBarrier

\section{Conclusion}
\label{sec:discussion}

WDF searches for gravitational wave transients without a template bank and
without assuming a waveform family at trigger generation. The strain is
conditioned by autoregressive whitening through a zero phase square root
filter, so that a transient's phase survives and every coefficient is in units
of one noise scale; each window is then transformed in several competing
orthonormal bases and the statistic is the norm of the coefficients that pass a
threshold fixed by the noise alone.
Two decisions shape what the pipeline delivers. The first is that events of
two detectors are admitted as a candidate when their supports are compatible
and not when their reported times agree: for an extended transient the reported
time is an estimator and it can differ between detectors by more than the light
travel time while the supports remain compatible, so the measured arrival time
difference is left to the ranking rather than spent on the gate. The second is
that selection is separate from characterization: the statistic is read on
the window the trigger was found in, the event's properties on the whole event
stitched from its triggers, which is what lets a signal longer than one
window be recovered and reconstructed.
The admissible inter detector pairs are finally ranked by a graph neural network defined on the coincidence graph. 
Single detector events form the nodes and physically admissible pairs form the edges. 
The ranking is therefore learned directly from the structure and attributes of the coincidence graph, 
rather than prescribed as an analytical combination of a few summary statistics. 
Since this graph is already constructed to identify admissible coincidences, 
the GNN does not require any modification of the upstream single detector search.
The model is trained on simulated data, selected using an independent held-out
realization and then applied without further tuning to real O4 strain, where it
ranks the same admitted pairs against the same background as the fixed formula
does. Section~\ref{sec:network-results} reports what each recovers, class by
class and rate by rate.

These numbers validate the pipeline; they are not a measurement of what the
search would deliver on an observing run. They rest on days of recorded strain
and on an injected population we chose, not on an astrophysical one, and the
rates they are read at are the rates that livetime resolves. Longer stretches
of real data are being analyzed, and the efficiency of the search at the rates
an observing run quotes will be reported there.

The main limitation concerns long signals and it is a severe one. A binary
neutron star stays in band for of order a hundred seconds and deposits, in
any window of a fraction of a second, an energy that thresholding removes. At
ten false alarms per day on real O4 strain the network recovers $0.11$ of them
against $0.95$ of the binary black holes. The cause is the window and not the ranking. An excess power search over fixed
windows cannot accumulate a signal to noise ratio spread over a thousand
windows, because accumulating it requires knowing along which track to sum, and
specifying that track requires a model.
On real O4 strain the pipeline behaves as it does on the simulated set and it
was also read on a real event. GW250114 is a detection and not an injection,
and with no waveform template the search returns it as the loudest zero lag
candidate of its segment; in both detectors the reconstructed energy centroid
precedes the reported geocentric peak time by less than the light travel time
from the geocenter to that detector. The segment is far too short for a
background to be measured on it, so this is a test of the machinery on a signal
nobody injected and not a statement about the event.

Conditioning and search cost about
$10$~ms per second of analyzed data for the two detectors together, on a
single core of a $3.0$~GHz AMD EPYC 7313, and the conditioning reads ahead of
what it emits by about three seconds, an amount the filters fix in advance.
Once a candidate exists, its reconstruction, its arrival time difference, its
GNN score and its sky ring are a few tens of milliseconds more, on that
core and on one NVIDIA RTX A5000 and that cost grows linearly with the event
rate.
The search is therefore compatible with a low latency analysis running in real
time and the latency of the chain is set by the front end and not by what
follows it. Each trigger carries the coefficients that produced it, the basis that
represented the window most compactly, the reconstruction and the noise scale.
That is a description of the transient which a classification chain can consume
without recomputing anything and it will be used in this way in Wavefier
\cite{wavefier}, which takes WDF as its trigger generator.  
What the search delivers, on the data analyzed here, is one workflow that
carries a transient from conditioned strain to a calibrated multi detector
candidate, a time domain reconstruction and a sky region, with no waveform
model entering any selection. Within the limit that long signals set, a
sparse wavelet representation supports detection, ranking and reconstruction
together.

Two directions are open and neither is settled here. The first is parameter
estimation by normalizing flow on the same coefficients
\cite{green2020dingo,wong2023jim,polanska2024flows}, which is the reason for
the pipeline's name. The second is a complete low latency test of the chain,
with the front end expressed on hardware such as FPGAs. The design decisions reported here
were taken so as not to preclude either.

\section*{Software and data availability}

\texttt{p4TSA}, the C++ core and its Python interface \texttt{pytsa}, is
available at \url{https://github.com/elenacuoco/p4TSA} and documented at
\url{https://p4tsa.readthedocs.io}. The pipeline \texttt{wdflow} is available
at \url{https://github.com/elenacuoco/wdflow} and documented at
\url{https://elenacuoco.github.io/wdflow/}. The simulated data set is generated by
\texttt{wdf.mock} and is reproducible from its seed.
\begin{acknowledgments}

The author gratefully acknowledges Alberto Iess, Francesco Di Renzo
 and Alessandro Staniscia for their valuable contributions throughout 
 the development of the WDF pipeline. Their feedback and technical support played an important role in improving
  the robustness and reliability of the implementation.
The author also thanks Edoardo Milotti for his careful reading of the
manuscript and for the suggestions that improved the paper.

Part of this work was carried out on BETIF, the computing and
analysis platform for the Einstein Telescope community operated jointly by the
Istituto Nazionale di Fisica Nucleare and the Department of Physics and
Astronomy of the University of Bologna,
\url{https://betif-difaet.readthedocs.io}.

This research has made use of data or software obtained from the Gravitational
Wave Open Science Center (gwosc.org), a service of the LIGO Scientific
Collaboration, the Virgo Collaboration and KAGRA. This material is based upon
work supported by NSF's LIGO Laboratory which is a major facility fully funded
by the National Science Foundation, as well as the Science and Technology
Facilities Council (STFC) of the United Kingdom, the Max-Planck-Society (MPS)
and the State of Niedersachsen/Germany for support of the construction of
Advanced LIGO and construction and operation of the GEO600 detector. Additional
support for Advanced LIGO was provided by the Australian Research Council. Virgo
is funded, through the European Gravitational Observatory (EGO), by the French
Centre National de Recherche Scientifique (CNRS), the Italian Istituto Nazionale
di Fisica Nucleare (INFN) and the Dutch Nikhef, with contributions by
institutions from Belgium, Germany, Greece, Hungary, Ireland, Japan, Monaco,
Poland, Portugal, Spain. KAGRA is supported by Ministry of Education, Culture,
Sports, Science and Technology (MEXT), Japan Society for the Promotion of
Science (JSPS) in Japan; National Research Foundation (NRF) and Ministry of
Science and ICT (MSIT) in Korea; Academia Sinica (AS) and National Science and
Technology Council (NSTC) in Taiwan.

The strain analyzed here comes from the second part of the fourth observing
run \cite{gwosc_o4b}.

The author acknowledges the use of generative AI in the preparation of this
manuscript: Claude Opus 5 (Anthropic), run through Claude Code, and ChatGPT
5.6 (OpenAI). They were used for language editing, for \LaTeX\ markup, for
drafting bibliographic entries whose details were taken from Crossref and
arXiv, and for reading the source code of \texttt{p4TSA} and \texttt{wdflow}
to check the statements this manuscript makes about what that code computes.
They were not used to design the method, write the pipeline or produce any
measurement reported here. The author directed and reviewed every change, keeps a record of
the use, and takes full responsibility for the content. This follows the
University of Bologna's guidelines on the use of generative AI in scientific
production.
\end{acknowledgments}

\appendix*

\section{Definitions and constructions}
\label{app:definitions}

\subsection{The zero phase whitening filter}
\label{app:conditioning}

The coefficients $a_k$ of Eq.~\eqref{eq:ar-noise-model} are estimated with
Burg's algorithm \cite{burg1975} on the $300$~s that open each analyzed
segment, and held fixed while that segment is processed, so the segment length
is what sets how closely the model tracks a drifting noise floor. No injection
is placed in that stretch: a transient inside it would be absorbed into the
estimated spectrum and whitened away with it rather than left for the search to
find. The filter is applied as a lattice recursion, which makes the
conditioning an online time domain operation rather than a sequence of spectral
estimates \cite{cuoco2001online,p4tsa}.

The filter applied in the two directions cannot be $A$ itself. Applied twice,
$A$ would weight the data by $1/S_x(f)$ rather than by $1/\sqrt{S_x(f)}$, and
the squared coefficient norm would scale as $1/S_x^2(f)$ instead of the
required $1/S_x(f)$. The filter to run in both directions is therefore the
one defined by Eq.~\eqref{eq:square-root-filter}.

$A_{1/2}$ comes from the model already available: a Levinson fit of order $q$
to the autocorrelation of the pseudo spectrum $1/\left|A(f)\right|$, once,
after the noise model is estimated and outside the processing loop. It is
therefore a finite impulse response filter of order $q$, whose reflection
coefficients the same lattice implementation takes \cite{p4tsa}. If $e_q$ is
the prediction error of that fit, the spectral factor satisfies

\begin{equation}
    \left|A_{1/2}(f)\right|^2
    \simeq
    e_q\left|A(f)\right|.
    \label{eq:square-root-approximation}
\end{equation}

Applying $A_{1/2}$ first forward and then backward gives

\begin{equation}
    \widetilde{y}(f)
    =
    e_q\left|A(f)\right|\widetilde{x}(f),
    \qquad
    S_y(f)
    =
    e_q^2\sigma^2,
    \label{eq:conditioned-spectrum}
\end{equation}

where $\widetilde{x}$ and $\widetilde{y}$ are the Fourier transforms of the
input and conditioned data: multiplying $S_x = \sigma^2/\lvert A\rvert^2$ by
$e_q^2\lvert A\rvert^2$ leaves the constant $e_q^2\sigma^2$. The conditioning
changes the shape of the spectrum and not the scale $\sigma$ of the noise
model, so the output is white, at zero phase, with standard deviation
$e_q\sigma$. We record that scale with the conditioned segment rather than
normalizing it away, so that an amplitude measured downstream can be returned
to the strain the detector recorded.

\subsection{The statistic as a matched filter signal to noise ratio}
\label{app:matched-filter}

Hard thresholding keeps a coefficient with its own amplitude or discards it,
so the reconstruction $\hat{\boldsymbol{h}}^{(b)}$ of
Eq.~\eqref{eq:wavelet-reconstruction} is the orthogonal projection of the
data onto the selected coefficient support. The data and the reconstruction
therefore satisfy

\begin{equation}
    \left\langle
    \boldsymbol{x},
    \hat{\boldsymbol{h}}^{(b)}
    \right\rangle
    =
    \left\lVert
    \hat{\boldsymbol{h}}^{(b)}
    \right\rVert_2^2 ,
    \label{eq:data-reconstruction-product}
\end{equation}

and the matched filter signal to noise ratio obtained by using the
reconstruction itself as a template is

\begin{equation}
    \frac{
    \left\langle
    \boldsymbol{x},
    \hat{\boldsymbol{h}}^{(b)}
    \right\rangle
    }{
    \sigma
    \left\lVert
    \hat{\boldsymbol{h}}^{(b)}
    \right\rVert_2
    }
    =
    \frac{
    \left\lVert
    \hat{\boldsymbol{h}}^{(b)}
    \right\rVert_2
    }{\sigma}
    =
    \EnWDF^{(b)} ,
    \label{eq:wdf-matched-filter-snr}
\end{equation}

which is Eq.~\eqref{eq:snr}. The caveats on this equality are stated in
Sec.~\ref{sec:statistic}.

\subsection{What an event is}
\label{app:parameters}

\begingroup
\squeezetable
\begin{table}[!ht]
  \centering
  \caption{What a single detector event and a network event carry.}
  \label{tab:pe}
  \begin{ruledtabular}
    \begin{tabular}{lp{0.50\columnwidth}}
      Quantity & Definition \\
      \colrule
      \multicolumn{2}{@{}l}{\emph{single detector event: a connected set of tiles}} \\
      \texttt{gpsCentroid} & energy centroid of the event's tiles in time \\
      \texttt{gpsPeak} & center of the tile carrying the largest coefficient \\
      \texttt{gpsEnvelope} & instant read on the event's stitched reconstruction \\
      \texttt{tSpread} & spread of the energy about the centroid, tile widths
        included \\
      \texttt{duration} & extent of the event's tiles \\
      \texttt{duration90} & interval holding the central 90\,\% of the energy \\
      \texttt{freqMin}, \texttt{freqMax} & edges of the event's tiles \\
      \texttt{freqMean} & energy weighted frequency of the tiles, taken in
        $\log f$ \\
      \texttt{freqQ05}, \texttt{freqQ95} & band holding the central 90\,\% of
        the energy \\
      \texttt{snrPeak} & largest coefficient on the noise scale,
        $\max\lvert c_k \rvert / \sigma$ \\
      \texttt{EnWDF} & \EnWDF\ read on the event's stitched reconstruction:
        what the event is worth \\
      \texttt{EnWDF\_window} & the largest \EnWDF\ among the event's windows:
        what ranks it \\
      \colrule
      \multicolumn{2}{@{}l}{\emph{network event: a pair the geometry admits}} \\
      \texttt{gps\_candidate} & mean of the two events' instants \\
      \texttt{dt\_s} & difference of the two events' own instants \\
      \texttt{dt\_over\_tolerance} & that difference as a fraction of the
        tolerance the pair was admitted with \\
      \texttt{frequency\_overlap} & band the two events share \\
      \texttt{time\_overlap} & time support the two events share \\
      \texttt{wavegram\_similarity} & cosine between the two normalized
        wavegrams \\
      \texttt{network\_enwdf} & the two events' loudness in quadrature \\
      \texttt{network\_min\_enwdf} & the smaller of the two \\
      \texttt{network\_morphology} & $R^{\mathrm{mor}}$ of Eq.~\eqref{eq:rank-morphology}:
        magnitude of the signed product over the tiles the two events share \\
    \end{tabular}
  \end{ruledtabular}
\end{table}
\endgroup

Table~\ref{tab:pe} lists what the search records at each of the two levels.
Every quantity of a single detector event is a moment over the event's own
tiles, on the noise scale $\sigma$ the search measured; every quantity of a
network event is a comparison of the two events it joins.

\FloatBarrier

\bibliographystyle{apsrev4-2}
\bibliography{references}

\end{document}